\documentclass[12pt]{article}
\usepackage{fullpage}

\usepackage[margin=1.0in]{geometry}
\usepackage{graphicx}
\usepackage{caption}
\usepackage{subcaption}
\usepackage[none]{hyphenat}
\usepackage{amsmath}
\usepackage{epstopdf}
\usepackage{natbib}
\usepackage{setspace}
\usepackage{subcaption}
\usepackage{wrapfig}
\newcommand{\MnervFull}{\emph{Megalonaias nervosa}}
\newcommand{\Mnerv}{\emph{M. nervosa}}
\newcommand{\Unio}{Unionidae}
\newcommand{\Rebekah}{Rebek\mbox{}ah }
\newcommand{\Cgig}{\emph{C. gigas}}
\newcommand{\CgigFull}{\emph{Crassotrea gigas}}
\newcommand{\Velli}{\emph{V. ellipsiformis}}
\newcommand{\VelliFull}{\emph{Venustaconcha ellipsiformis}}
\newcommand{\Ehope}{\emph{E. hopetonensis}}
\newcommand{\EhopeFull}{\emph{Elliptio hopetonensis}}
\newcommand{\Ecras}{\emph{E. crassidens}}
\newcommand{\EcrasFull}{\emph{Elliptio crassidens}}

\newcommand{\Ecomp}{\emph{E. complanata}}
\newcommand{\EcompFull}{\emph{Elliptio complanata}}

\newcommand{\Dros}{\emph{Drosophila}}
\newcommand{\Lcar}{\emph{L. cardium}}

\newcommand{\Llien}{\emph{L. lienosus}}
\newcommand{\LlienFullBoth}{\emph{Leaunio lienosus} (\emph{Vilosa lienosa})}

\newcommand{\Aplic}{\emph{A. plicata}}
\newcommand{\AplicFull}{\emph{Amblema plicata}}
\newcommand{\Utet}{\emph{U. tetralasmus}}
\newcommand{\UtetFull}{\emph{Uniomerus tetralasmus}}
\usepackage[ocgcolorlinks,citecolor=blue]{hyperref}
\usepackage{lineno}
\begin{document}

\author{ \small \Rebekah L. Rogers$^{1*}$, Stephanie L. Grizzard$^{1,2}$, James E. Titus-McQuillan$^{1}$, Katherine Bockrath$^{3,4}$\\
\small Sagar Patel$^{1,5,6}$, John P. Wares$^{3,7}$, Jeffrey T Garner$^{8}$, Cathy C. Moore$^{1}$}

\title{Gene family amplification facilitates adaptation \\in freshwater unionid bivalve \MnervFull}
\date{}
\maketitle

\noindent Author Affiliations: \\
1.  Department of Bioinformatics and Genomics, University of North Carolina, Charlotte, NC \\
2.  Department of Biological Sciences, Old Dominion University, Norfolk, VA \\
3.  Department of Genetics, University of Georgia, Athens, GA \\
4.  U.S. Fish and Wildlife Service, Midwest Fisheries Center Whitney Genetics Lab, Onalaska, WI\\
5.  Department of Biology, Saint Louis University, St. Louis, MO \\
6. Donald Danforth Plant Science Center, St. Louis, MO \\
7. Odum School of Ecology, University of Georgia, Athens, GA \\
8. Division of Wildlife and Freshwater Fisheries, Alabama Department of Conservation and Natural Resources, Florence, AL \\

\vspace{0.5in}

\noindent *Corresponding author:  Department of Bioinformatics and Genomics, University of North Carolina, Charlotte, NC.  Rebekah.Rogers@uncc.edu \\

\vspace{0.5in}

\noindent Short Title: Gene family expansion in \MnervFull \\

\clearpage

\subsubsection*{Abstract}

Freshwater unionid bivalves currently face severe anthropogenic challenges.  Over 70\% of species in the United States are threatened, endangered or extinct due to pollution, damming of waterways, and overfishing.   These species are notable for their unusual life history strategy, parasite-host coevolution, and biparental mitochondria inheritance.  Among this clade, the washboard mussel \MnervFull {} is one species that remains prevalent across the Southeastern United States, with robust population sizes.  We have created a reference genome for \Mnerv {} to determine how genome content has evolved in the face of these widespread environmental challenges.  We observe dynamic changes in genome content, with a burst of recent transposable element proliferation causing a 382 Mb expansion in genome content.  Birth-death models suggest rapid expansions among gene families, with a mutation rate of $1.16 \times 10^{-8}$ duplications per gene per generation.  \emph{Cytochrome P450} gene families have experienced exceptional recent amplification beyond expectations based on genome wide birth-death processes.  These genes are associated with increased rates of amino acid changes, a signature of selection driving evolution of detox genes.  Fitting evolutionary models of adaptation from standing genetic variation, we can compare adaptive potential across species and mutation types. The large population size in \Mnerv {} suggests a 4.7 fold advantage in the ability to adapt from standing genetic variation compared with a low diversity endemic \Ehope. Estimates suggest that gene family evolution may offer an exceptional substrate of genetic variation in \Mnerv, with $P_{sgv}=0.185$ compared with $P_{sgv}=0.067$ for single nucleotide changes.  Hence, we suggest that gene family evolution is a source of ``hopeful monsters" within the genome that may facilitate adaptation when selective pressures shift.  These results suggest that gene family expansion is a key driver of adaptive evolution in this key species of freshwater \Unio {} that is currently facing widespread environmental challenges.   This work has clear implications for conservation genomics on freshwater bivalves as well as evolutionary theory.  This genome represents a first step to facilitate reverse ecological genomics in \Unio {} and identify the genetic underpinnings of phenotypic diversity.  

\clearpage
\subsubsection*{Introduction}

As organisms are faced with intense rapidly changing selective pressures, new genetic material is required to facilitate adaptation.  Among  sources of genetic novelty, gene duplications and transposable elements (TEs) offer new genes or new regulatory patterns that can facilitate evolutionary change \citep{oliver2009,Feschotte2008,conant2008,Ohno1970}.  These mutations can offer a substrate of novel genetic variation that may produce new functions to help organisms adapt when innovative changes are needed for survival.    In evolutionary systems that are faced with sudden shifts in selective pressures, these sources of new genes are expected to be all the more essential as large-effect mutations facilitate moves to novel fitness peaks \citep{Orr2005}. 

 Historically, duplicate genes and TEs have been well studied in model organisms where it has been possible to estimate prevalence \citep{Adams2000,12Genomes, Lander2001, goffeau1996}, mutation rates \citep{Bennett2004, Adrion2017, Demuth2006,TripleHan,Rogers2009}, and identify adaptive changes \citep{aminetzach2005,Han2009,schmidt2010}.  With advances in genome sequencing it is possible to gain a broader view of how gene family proliferation and TE content evolve across the tree of life.  Most model organisms are either short lived with large, persistent populations (\Dros, yeast) or long lived with modest population sizes and lower genetic diversity (Humans, mammals) \citep{LynchBook}.  It has only recently become tractable to generate reference genomes and evaluate the role of genetic novelty in alternative evolutionary systems.  An understanding of how genomes evolve in populations that have recently faced strong shifts in selective pressures and sharp population declines that amplify genetic drift will clarify whether these mutations might play an outsized role in specific evolutionary scenarios.  
 
The tempo of evolution in response to selection depends directly on population sizes, mutation rates, and generation time \citep{Hermisson2005,Gillespie1991,Maynard1971}.  It is therefore important to understand how genetic novelty responds in systems with alternative demography and life history in order to discern how these loci contribute to rapid evolutionary change. Moreover, the potential for species to respond to selective challenges may depend on life history, as maturation time, growth and reproductive strategies may contribute to risk of extinction \citep{haag2011,haag2012,Stearns2000,Johnson2002}.    Freshwater unionid bivalves (\Unio) offer a long-lived evolutionary model \citep{haag2011,haag2012}, with intense changes in population sizes and strong selective pressures due to environmental change \citep{Strayer2004,Haag2014}.  These bivalves have experienced systemic decline due to human-mediated habitat alteration to freshwater ecosystems, with approximately 70\% of North American species considered threatened, endangered, or extinct \citep{williams1993,Strayer2004,Haag2014}.  Understanding of how duplications and TEs have evolved as these taxa face modern anthropogenic challenges may explain how they might contribute to species survival during sudden evolutionary changes.

Unionid bivalves are benthic filter feeders that form the backbone of freshwater ecosystems \citep{Lopez2020, williams2008,Vaughn2008,Vaughn2018}.  They are notable for their unusual life history and mitochondrial biology. \Unio {} are known for biparental mitochondria inheritance, where females receive mitochondria solely from the maternal lineage but males also receive germline mitochondria from paternal lineages, a potential genetic factor in sex determination \citep{Breton2007, Breton2011, Liu1996, Wen2017}.  While adults are sessile filter feeders, all species have a larval stage that parasitizes fish gills, facilitating dispersal \citep{Barnhart2008, williams2008}.  Because of this unusual life history, \Unio {} serve as an interesting evolutionary model system for parental care, parasite-host coevolution, and evolution of life history strategies.   It may also offer a system to study the genetic response to strong and suddenly shifting selective pressures \citep{Lopez2020, williams2008} in a long-lived species \citep{haag2011, haag2012} that departs from the standard population genetic models.


Given the magnitude of shift in freshwater ecosystems, \Unio {} are an ideal model to study how new gene formation creates ``hopeful monsters" within the genome and facilitates rapid adaptive change.  Genome sequencing offers a means to bring these fascinating organisms into the modern genetic era, where reverse ecological genomics may reveal the genetic responses to these changing environmental and coevolutionary pressures \citep{marmeisse2013}. Many studies have surveyed genetic diversity using a small number of loci, microsatellites, mtDNA, or more recently transcriptomes or SNP capture arrays \citep{Campbell2005, Pfeiffer2019,Cornman2014,Gonzalez2015,Wang2012}.  However, full genome sequencing is not yet widespread, limiting genetic analysis.  As a first step to facilitate evolutionary genetics with sufficient power to resolve genetic patterns of adaptation in \Unio, sequencing and assembly of reference genomes can help fill this gap.  Advances in genome annotation and assembly will have broad impacts on conservation and management for imperiled \Unio.  

Among \Unio, \MnervFull {} is one widespread, prevalent species whose populations have remained robust, so far, in the face of environmental challenges.  \MnervFull {} is thick-shelled, spawns in the fall, and produces large brood sizes of up to 1 million glochidia per individual \citep{williams2008, Haggerty2005}.  Earliest maturation is between 5-8 years of age \citep{woody1993} and the species is long lived with individuals identified up to 43 years of age \citep{haag2011}. \Mnerv {} larvae are generalist parasites that use multiple prevalent fish species as hosts \citep{woody1993, williams2008}.  These  mussels are present throughout the Southeastern United States, in the Mississippi, Ohio, and Tennessee River waterways \citep{williams2008}. The Tennessee River offers a focal location where 79 species of Unionidae have been reported, though as many as 32 of those are extirpated or extinct \citep{Garner2001}.  In Northern Alabama, The Tennessee Valley Authority (TVA) has documented as much as 6,700 tons of shells harvested in a single year at peak demand \citep{williams2008} and up to 45\% of these were \Mnerv {} \citep{Ahlstedt1992}.   

To determine how the genome of \Mnerv {} has responded to recent selective pressures, we have generated a reference genome for a single individual collected from Pickwick Lake, Muscle Shoals, Alabama.  We explore the relative contribution of gene family expansion, TE proliferation, and single nucleotide mutations to genetic variation and its influence on evolutionary potential.   This reference genome shows strong signatures of TE proliferation and adaptation through gene family expansion in \Mnerv.  We further analyze genetic diversity in comparison with recently sequenced bivalves, and how it is expected to influence subsequent evolutionary processes in this species.  

\subsubsection*{Methods}

\subsubsection*{Assembly and annotation}
A single gravid female roughly 11 years of age based on ring counts was collected from Pickwick Lake in Muscle Shoals, Alabama, in 2017 (JGPickwickLake2017 Specimen \#3).  The individual was 15.8 cm long (anterior to posterior), 11 cm high (dorsal to ventral), and 5.7 cm wide (between outer surfaces of the two shell valves), with large growth rings identified in the earliest years. The individual was shipped to Charlotte, North Carolina, where it was dissected and tissues were flash frozen in liquid nitrogen.  We extracted DNA from compact adductor tissue using the Qiagen DNeasy kit.  Illumina TruSeq DNA PCR-free libraries were prepared manually following the manufacture's protocol targeting 350bp inserts. Size selection was optimized using Ampure XP beads. The libraries were sequenced (2 lanes) on the Illumina HiSeq 4000 platform using the 150bp paired end (PE) Cluster kit (DUGSIM, Duke University). To obtain ultra-long reads, 1D libraries were prepared from tissue samples and sequenced on Oxford Nanopore (DeNovoGenomics).  High molecular weight DNA was extracted using a phenol-chloroform preparation and purified using pulse-field gel.  An attempt to scaffold the reference individual using Hi-C sequencing at the NC State sequencing core was unsuccessful, though it is unclear whether this was due to tissue integrity or technical problems.  DNA purification from molluscs is often challenging because of polysaccharides and polyphenols \citep{Pereira2011}, a factor that may require optimization in future experimental design.

Sequences were assembled using the Masurca hybrid assembler \citep{masurca}.  We used BUSCO 3.0.2 \citep{waterhouse2017, simao2015} to assess reference quality and completeness using the metazoan ortholog set.  Some assembled contigs reflect algae and bacterial contamination, as expected.  These contigs were identified using a BLASTn (v. 2.5.0+) search against the algae (JGI), trematode, schistostomate  and bacterial databases (NCBI) and excluded from the final assembly and annotations.   The female mitochondrial contig, jcf7180000235847, was identified using a BLASTn search against the \emph{Quadrula quadrula} mtDNA sequence obtained from NCBI (accession NC\_013658.1, downloaded Aug 15 2018).  

In addition we obtained transcriptomes from 6 tissues: glochidia, gills, mucus glands, mantle, adductor muscle, and muscular foot. Tissues were flash frozen in liquid nitrogen immediately on dissection, then RNA was extracted using Zymo DirectZol RNA microPreps without DNase.  RNA was converted to cDNA using the Illumina TruSeq mRNA HT Library Prep.  Libraries were prepared using 8M fragmentation (Covaris) and 15 cycles of amplification. We used secondary cleanup step with Ampure XP beads to remove lower molecular weight primer-dimers. Transcriptomes were sequenced on a single lane of a HiSeq4000 using the 150bp paired end (PE) Cluster kit (DUGSIM). 

We assembled transcriptomes using the Oyster River Protocol v2.3.1 (ORP) \citep{ORP}, then aligned to the reference genome using BLAT v. 36 \citep{kent2002}.  Alignments were converted to hints using blat2hints and filtered for unique matches using filterPSL from Augustus. ORP is able to assemble transcripts with TPM$<1.0$, better performance than previous \emph{de novo} assemblers including Trinity \citep{ORP}.  Their OrthoFuser program combines transcripts that are homologous, again with better performance than any single assembler alone \citep{ORP}.  These alignment hints were used as input for \emph{de novo} gene prediction in Augustus v3.3 \citep{stanke2003}, allowing isoform prediction with evidence \citep{stanke2008}.  Out of 709,458 putative transcripts identified across 6 tissues, 641,750 were aligned to a single contig.  Species-specific gene prediction models for \Mnerv {} produced during BUSCO training were used for Augustus gene prediction.  We used Augustus gene models to create CDS and protein fasta files using gffread \citep{gffread}.  We produced functional annotation for gene models using Interproscan v5.34-73.0 \citep{quevillon2005} based on identified Pfam \citep{pfam} conserved domains.  

Annotated sequences were compared to the RepBase database \citep{jurka2005} of TEs in a BLASTn search at an E-value of $10^{-10}$.  Annotations with BLAST hits in RepBase or with Interproscan results suggesting TE or viral activities were removed from the annotations.  We note that these criteria could allow for some novel selfish genetic elements if sufficiently distant from previously annotated TEs in other species.  Sequences were compared against gene models from \CgigFull {} (the Pacific Oyster) \citep{oyster2012} using a BLASTp at an E-value of $10^{-20}$.  One-to-one orthologs were defined as reciprocal best hits.  To establish a final high quality annotation data set, we required that \emph{de novo} predictions be supported by at least one line of evidence: results from Interproscan, BLAST comparison against \Cgig {} or support in the ORP transcriptome assembly.   

Differences in gene annotations could potentially be caused by methodological differences in studies on various species.  To generate putative annotations that are directly comparable, we generate transcriptome based annotations for six species of \Unio: \Mnerv, \Utet, \Ecomp, \Ecras, \Llien, and \Velli {} see Supporting Information).

\subsubsection*{Gene Families}
Gene family expansion is an important mode of new gene formation that can lead to evolutionary innovation \citep{TripleHan,conant2008,Hahn2009,Ohno1970}.  The rates at which new genes are formed and lost dictates evolution of gene content and the substrate of genetic novelty \citep{Lynch2000, Rogers2009,Hahn2009}. To determine the genomic responses of new gene formation in \Mnerv, we identified gene families and used birth-death models to identify rates of evolution through new gene formation \citep{Lynch2000,Rogers2009}.  These models assume new genes form in clock-like fashion according to Poisson processes.  The history of gene birth and death is recorded in the demographic history of gene family members, where gene ages are recorded in genetic sequence data.  These methods offer a more accurate estimate of new gene formation than simple accounting where the number of new genes observed is divided by branch lengths \citep{Lynch2000}.  They are also tractable to study birth-death processes for duplicate genes in a single genome.

We used a Fuzzy Reciprocal Best Hit algorithm \citep{TripleHan} on an all-by-all BLASTn comparison at an E-value of $10^{-20}$ of gene annotations within \Mnerv {} to identify gene families. We allowed for a difference in E-values within two orders of magnitude.   If orthogroup order was ambiguous, the maximum rank was chosen.  The distribution of duplicate gene ages is recorded in the divergence among copies at synonymous sites (dS) and the relative rate of amino acid changing mutations is recorded in divergence at non-synonymous sites (dN). 

We collected sequences for each orthogroup, translated them, then aligned amino acid sequences in clustalw2 \citep{clustalw, clustalwx}.  Alignments were back-translated to the original nucleotide sequence to produce in-frame alignments.  We used PAML's codeml \citep{PAML} to estimate dN and dS, estimating codon frequencies from nucleotide frequencies (F1x4) under the guide tree from clustalw2. We excluded all genes past the mutation saturation point with $dS >1.0$.  We then fit a birth-death model using maximum likelihood to the distribution of dS for gene family members according to $\lambda e^{-\mu dS}$ similar to previous work on duplicate genes \citep{Lynch2000,Rogers2009}.   We established the 95\% CI using a jackknife of 75\% resampling with 10,000 replicates.  Mutation rates for duplicate gene formation and the half-life of mutations were inferred based on birth-death parameters.  These can be rescaled to years assuming copies diverge by twice the mutation rate of $5\times 10^{-9}$ \citep{scallop}.  Interproscan conserved domain annotations identified orthogroups belonging to large gene families. We fitted separate birth-death models for the most common gene families in the \Mnerv {} genome: \emph{Cytochrome P450} genes, \emph{Hsp70s}, \emph{Mitochondria-eating} genes, Chitin metabolism genes, \emph{von Willebrand} proteins, \emph{ABC transporters} and \emph{opsin} genes.

\subsubsection*{TE content}
Transposable elements are known as sources of innovation that can remodel genomes to produce innovation  \citep{oliver2009,Feschotte2008}.  They are often associated with genome remodeling not solely because of new insertions but also due to their ability to facilitate rearrangements and copy gene sequences.  To determine how TEs might have reshaped genome content in \Mnerv, we identified putative transposable elements in \Mnerv {} using a three pronged approach.   Repeatscout 1.0.5 \citep{RepeatScout}, RepDeNovo \citep{Chu2016}, and homology-based detection using either Interproscan 5.34-73.0 \citep{quevillon2005} or BLAST comparisons to RepBase \citep{RepBase}.  RepeatScout identifies repetitive sequences in genomes where TE copies have sufficient divergence from one another to successfully assemble into independent contigs.  We then mapped these repeats back to the \Mnerv {} reference assembly using a BLASTn at an E-value of $10^{-20}$ with dust filters turned off.  These methods effectively identify repeats with greater than 2\% divergence.  They do depend on genome assembly quality, and may not be appropriate for cross-species comparisons when assembly quality is variable. 

RepDeNovo does not rely on pre-existing genome assemblies but rather assembles contigs from high copy number reads in raw fastq files.  This program identifies recently proliferated transposons that have less than 2\% divergence from one another.  We mapped Illumina sequencing reads onto these repetitive contigs identified with RepDeNovo and estimated copy number using sequence depth.  TE types were identified using a tBLASTx against the RepBase Database.  Finally, we identified transposon proteins among gene annotations using homology based searches in Interproscan or using BLASTn comparisons with the RepBase database. Together, these methods generate a comprehensive portrait of TE content in the genome of \Mnerv.   

For comparison, Illumina sequences are incidentally available for \VelliFull {} (42X coverage, SRR6689532) \citep{Renaut2018} and \EhopeFull {} (14X coverage), both originally collected for other purposes.  \VelliFull {} is present throughout the northern Mississippi River, while \Ehope {} is an endemic species with limited numbers located in the Altahama River.  These distantly related taxa offer the nearest comparisons among unionid mussels where genetic data are available.  We used the assembly-free methods of RepDeNovo \citep{Chu2016} to assay TE content in these two species.  These data offer directly comparable estimates of TE content, independently from quality of genome assembly.  We used downsampled data for \Mnerv {} at 14X coverage to ascertain that differences in TE content were not driven by sequence coverage.  

\subsubsection*{Low coverage assembly for \Ehope}
Illumina sequencing data were incidentally available for \EhopeFull {} taken from a single individual collected from the Ocmulgee River in 2012 (Field ID JMW120329-2, Ocmulgee R, site 2 of day, JMW, MJH, DAW).  Mantle tissues was dissected and prepared for DNA sequencing on a HiSeq2000 at the University of Colorado sequencing core.  Using this 14X coverage Illumina sequencing data for \Ehope, we have generated a cursory reference assembly in Abyss using default parameters.  The assembly N50 is only 1 kb, and the maximum contig size is 22 kb.  Genome size estimates are 1 Gb, with 500 Mb in scaffolds 1kb or larger.   BUSCO \citep{waterhouse2017, simao2015} analysis suggests the assembly is largely incomplete with only 15\% of conserved metazoan genes identified.  However, in a tBLASTn, we can find partial matches for 15,010 \Cgig {} genes.  The number of \Cgig {} orthologs identified in this cursory reference is similar to the number identified in \Mnerv.  While this cursory reference genome is not of sufficient quality to anchor genes to regions or create syntenic maps, it can offer a comparison against \Mnerv {} for genome size, and effective population size, as well as surveys of TE content described above. 

\subsubsection*{Effective population size}

The tempo of evolution depends directly on the effective population size, mutation rate, and generation times.  To evaluate the timescale of adaptation, we used heterozygosity in the reference specimen to estimate population genetic parameters such as $\theta=4N_e\mu$, the expected coalescent time for neutral alleles of 2N generations, and $t_{MRCA}$ of 4N generations  \citep{Wakeley2009}.  To further identify the genetic potential available through standing genetic variation, we estimated the time to establishment of selective sweeps on new mutations $T_e$, and the probability of adaptation from standing variation $P_{sgv}$.  These parameters offer a description of theoretical expectations for how evolution should proceed given current population sizes and estimates of diversity.  They offer a means to determine the limits of what might be possible given strong selection acting on these species. 
 
 We mapped Illumina reads from the genome assembly to the \Mnerv {} reference genome using bwa aln with default parameters \citep{BWA}.  We identified SNPs using GATK \citep{gatk} haplotype caller v. 3.8-0-ge9d806836 assuming diploid genomes.  We identified heterozygous sites to estimate genome wide heterozygosity for \Mnerv.  We performed identical analysis for the cursory assembly for \Ehope.  These data provide estimates of $\theta=4N_e\mu$ for these freshwater bivalves.  Using these data we calculate $N_e$ based on a mutation rate of $\mu=5 \times 10^{-9}$ in bivalves \citep{scallop}.  We are able to use this parameter to estimate coalescent timescale of $2N$ generations and the $t_{MRCA}=4N$ generations.  Timescales are offered in generations and converted to years assuming an average 20 year generation time based on the long lifespan and late maturation of \Mnerv {} at 8 years of age \citep{woody1993,haag2011}. Alternative scaling from generations to years would scale linearly with generation time.  

We estimate the time to establishment of a sweep on a new mutation $T_e=\frac{1}{\theta s}$ \citep{Gillespie1991,Maynard1971} for each species.  These models describe time for new mutations to arise \emph{de novo} and escape stochastic forces of drift to establish selective sweeps.  The alternative to adaptation from new mutation is adaptation from standing genetic variation.  Population diversity depends directly on the effective population size and mutation rate \citep{watterson1975}.   Yet many mutations present in populations will be at such low frequency that they are unable to establish deterministic selective sweeps in the face of genetic drift \citep{Hermisson2005}.  We estimated the probability of adaptive pressures are able to establish deterministic sweeps from standing variation $P_{sgv} = 1 - e^{-\theta*ln(1+2N_e s)}$ \citep{Hermisson2005}.   It depends on the population size, selective coefficient, and mutation rate.   These parameters are estimated assuming moderate selective coefficients of $s=0.01$ and for very strong selection of $s=0.1$. 
  
\subsubsection*{Gene trees and evolutionary relationships}  
We generated transcriptome based annotations for 6 species of \Unio:  \Mnerv, \Utet, \Ecomp, \Ecras, \Llien, and \Velli.  For each of 6 assembled transcriptomes, we identified 5285 sets of reciprocal best hit orthologs with \Mnerv {} across all transcriptomes. Sequences with a direct match across all 6 species were used to generate gene tree phylogenies and address Incomplete Lineage Sorting (ILS). We aligned sequences using MUSCLE \citep{Edgar2004} under default settings. Site trimming was conducted using trimAl 1.2 \citep{Capella2009} with a no gap penalty among alignments of the annotated \Mnerv {} gene. We chose to use a no gap penalty criterion to mitigate potential bioinformatics artifacts and exon boundary uncertainties among the available transcriptomes. A total of 4475 gene sets met these criteria. We used these alignments to generate gene tree phylogenies using RAxML 8.2.12 \citep{Stamatakis2014,Stamatakis2006}. The rapid bootstrapping function was selected using a GTR $\Gamma$ model across 100 randomly sampled tree replicates. Astral III \citep{Zhang2018} estimated Normalized Quartet Scores \citep{Sayyari2016} and constructed a consensus tree among all generated gene trees from RAxML. Tree topology was plotted in R package Phangorn \citep{Schliep2011} using the DensiTree function \citep{Bouckaert2010}, using unscaled branch lengths for the proposes of visualization. Full Astral III consensus tree use coalescent scaled branch lengths (t/2N units) and gene tree branch lengths in substitutions are reported in Supporting Information.   Gene trees from RAxML were used to identify discordance with the consensus phylogeny.

  
\subsubsection*{Results}

\subsubsection*{Genome Assembly}
 
We generated a reference genome sequence for \Mnerv {} to facilitate evolutionary and ecological genomics in \Unio.  We collected 40X Illumina short read sequencing and 5X Oxford Nanopore long molecule sequences for a single gravid female \Mnerv {} from Pickwick Lake, Muscle Shoals, Alabama. Genome statistics prior to filtering were 2.7 Gb, with a scaffold N50 of 52.7 kb and a contig N50 of 51.5 kb (Table \ref{GenomeStats}).  These show higher contiguity than the previously assembled genome of \Velli {} \citep{Renaut2018}.  We compared these results with one high quality marine bivalve assembly for Pacific oyster, \emph{Crassotrea gigas}.  Divergence between \Cgig {} and \Mnerv {} is difficult to estimate with certainty, but ranges from 495 million \citep{TimeTree} to 200 million years \citep{bolotov2017}.  While these marine and freshwater bivalves that diverged hundreds of millions of years ago are phenotypically and ecologically distinct, \Cgig {} is the nearest non-unionid bivalve with a high quality assembly and annotation. The contig N50 is superior to the recent \Cgig {} genome of  19.4 kb, but poorer than the scaffold N50 of 401 kb \citep{oyster2012}.  The distribution of kmers is continuous with no ``shoulders" of high or low kmer frequency, indicating no sign of polyploidy or genomes misassembled in the face of heterozygosity (Figure \ref{kmer}).  

Some 92\% of the genome is assembled into contigs over 10 kb and 50\% is assembled into contigs 50 kb or larger.  Contaminating sequences were removed from the bivalve reference assembly using a BLASTn, producing a total assembly of 2.36 Gb, 500 Mb (28\%) larger than estimates for one other unionid mussel sequenced to date, \Velli.  \VelliFull {} has a genome size of 1.80 Gb \citep{Renaut2018}. BUSCO analysis \citep{simao2015} suggests 83\% complete orthologs and 9\% fragmented orthologs, consistent with previous genome assemblies for bivalves.  Only 2.1\% of BUSCOs are duplicated, again confirming the absence of polyploidy or high heterozygosity that would lead to separation of homologous chromosomes in the assembly.  Such sequence quality captures the majority of gene content either wholly or partially, and is sufficient for many applications in evolutionary and population genomics.



\subsubsection*{Gene Annotations}

\begin{figure}
\begin{center}
\includegraphics[scale=0.3]{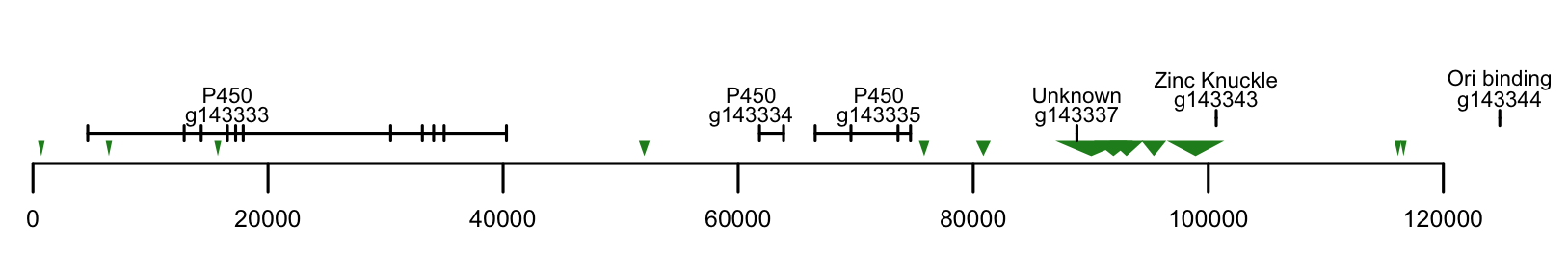}
\end{center}
\caption{A 127,552 bp contig containing 3 \emph{cytochrome P450} genes. Exons are represented with rectangles, introns with connecting lines.  Repetitive elements identified using RepeatScout are marked in green.  Genes on the reverse strand are marked below and genes on the forward strand are above. \label{annotation}}
\end{figure}

To identify genes and gene families that have emerged in freshwater \Unio, we generated gene annotations for \Mnerv.  We produced gene annotations using RNA sequencing data for 6 tissues from the same individual: Gills, mantle, muscular foot, mucosal palps, adductor muscle, and glochidia (offspring) (Figure \ref{annotation}).  The Oyster River Protocol identifies 66.7-87.1\% of BUSCOs across each of the 6 transcriptomes.  We identify  49,149 genes: 17,982 singleton genes with 21,596 transcripts and 31,167 members of gene families with 32,967 transcripts.  A total of 801 transcripts belonging to 733 genes contain premature stop codons suggesting pseudogenes.   Genes encompass approximately 477 kb of sequence.  Average intron length is 3714 bp, (min 7, max 145,991).  In a reciprocal best hit BLAST comparison, we identify 8811 one-to-one orthologs between \Mnerv {} and \Cgig.  

\begin{figure}
\begin{center}
\includegraphics[width=0.4\textwidth]{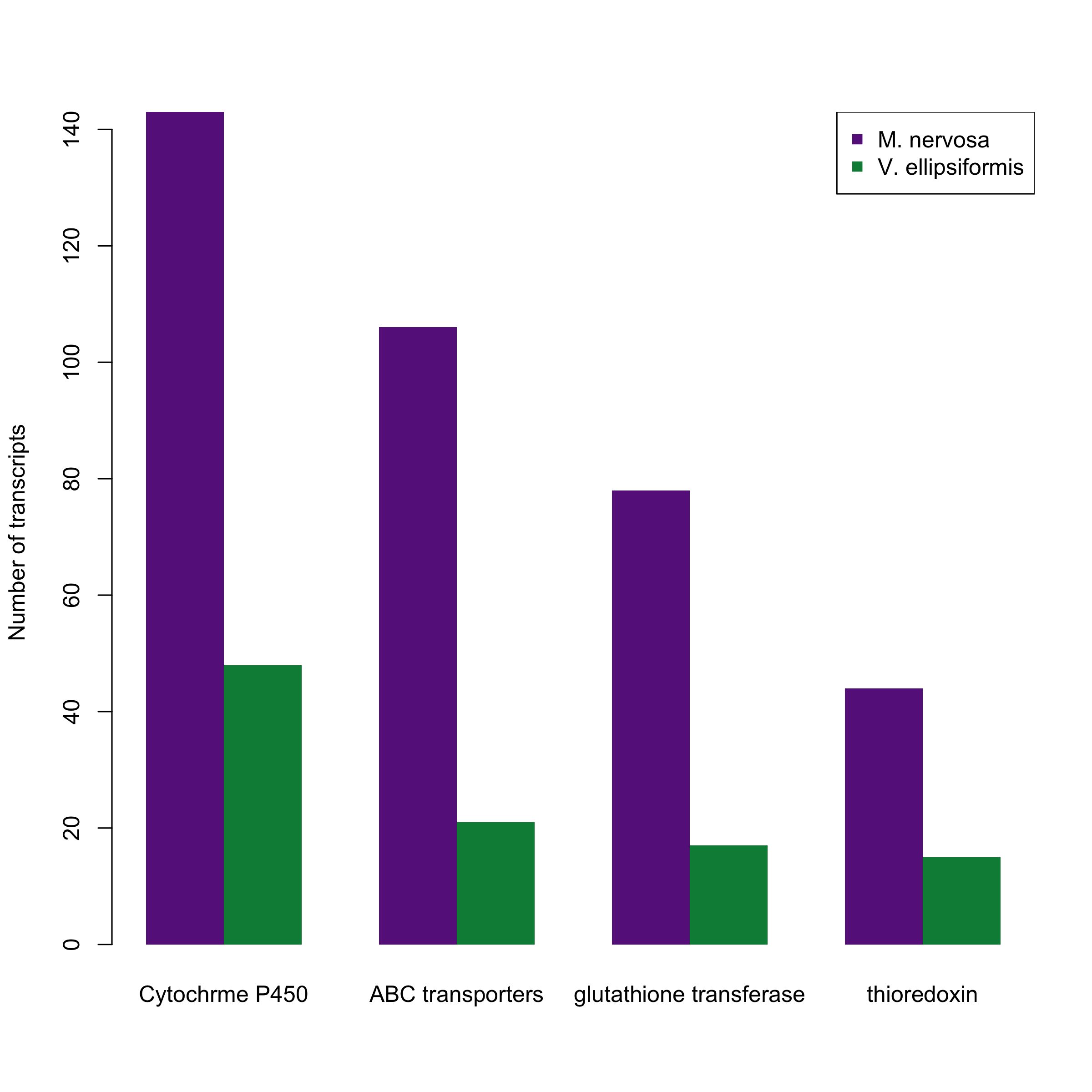}
 \caption{Detox genes among annotations for \Mnerv {} and \Velli. \label{detox}}
 \end{center}
 \end{figure}

We placed no \emph{a priori} criteria for what functional categories should be represented in the \Mnerv {} genome.  However, among Interproscan annotations \citep{quevillon2005}, we note large gene families related to biological processes expected to be important for \Unio {} survival and reproduction.   We identify 78 glutathione transferase genes, 44 thioredoxins, 143 cytochrome P450 genes, and 106 ABC transporters expected to interact with cytochromes (Figure \ref{detox}).  In BLAST searches we observe over 500 fragmented copies each of \emph{cytochrome P450} genes and \emph{ABC transporter} interacting regions that are not of sufficient length for identification in Interpro scans of functional domains.  Only one \emph{cytochrome P450} gene overlaps with another annotation, a gene of unknown function.  Thus, it is not likely that isoform splitting could be artificially inflating gene family sizes in the annotations.  We find multiple copies of genes involved in anticoagulation processes in \Mnerv, consistent with blood feeding during the parasitic larval stage. We find 138 von Willebrand factor proteins (99 type A and 39 type D), 15 Xa-binding  genes, and 33 fibrinogen binding proteins. 

We further observe 96 \emph{Hsp70}s and 6 \emph{Hsp90}s, consistent with previously published results from \Cgig {} showing expansion of heat shock proteins \citep{oyster2012}, but substantially increased over only 52 observed in \Velli {} \citep{Renaut2018}. These genetic differences are consistent with thermal tolerance through \emph{Hsp} activity in other organisms \citep{Sorensen2003}, though pleiotropic functions of \emph{Hsp70} genes may also influence gene family expansion. \Velli {} has a range in the northern US, cooler climates compared with \Mnerv {}.  \MnervFull {} is found throughout the Southeastern United States, and the present sample's home location lies in Muscle Shoals, Alabama.  We also note 238 rhodopsin genes and 117 Mitochondria-eating protein genes.  These are in addition to large numbers of conserved domains with less clear functional implications such as zinc finger domains, WD-40 domains, short chain dehydrogenase domains, Ankyrin repeats.  In a targeted identification, we find 11 chitin synthase genes and 55 Chitin binding Peritrophin-A genes, putatively important for shell formation.  We identify 7 mucins, proteins important for defense and release of glochidia offspring.  These numbers are many times higher than listed in official annotations in \Velli {} (Table \ref{CompVelli}), though we note that annotation methods differ, especially with respect to repeat masking before annotation rather than after annotation.   

Previously published annotations for \Velli {} show fewer members of these large gene families (Table \ref{CompVelli}) \citep{Renaut2018}.  However, these annotations through different methods and with different assembly qualities and different tissues could lead to a false portrait of species difference. To generate annotations for direct comparisons independently from genome assembly, we annotated \emph{de novo} assembled transcriptomes for \MnervFull, \EcrasFull, \EcompFull, \UtetFull, \AplicFull, \LlienFullBoth, and \VelliFull {} (Supplementary Information).   Results of transcriptome only analysis point to large detox gene content across \Unio, but with variation in the extent of gene family expansion (Figure \ref{detox2}, Table \ref{TransCypDetox}).  \MnervFull {} does not appear to be an outlier with respect to any functional category (Table \ref{detox2}, Figure \ref{TransCypDetox}).  These genome free transcriptome annotations validate results of large gene families from genome based annotations.  They also suggest that copy number expansion is common across \Unio, a prospect that deserves greater attention with whole genome annotation in the future.    

%
\subsubsection*{Gene Family Evolution}
Gene family expansion through duplication can offer a substrate of genetic novelty that may facilitate adaptation to new environments.  To identify signatures of adaptive changes and rapid evolution of copy number, we estimated $dN/dS$ \citep{Nei1986,Goldman1994} and fit birth-death models genome wide and for large gene families in \Mnerv.  We also evalutated putative tandem duplications that may be heterozygous in the reference genome alone by mapping Illumina reads to the reference and finding divergently oriented read pairs \citep{Rogers2014}.  We identify 8507 cases where abnormally mapping read pairs would suggest tandem duplications in one of the two haplotypes, in addition to those gene duplications identified in the reference assembly here, though not all of these are expected to capture gene sequences. 

 \emph{Cytochrome P450} family orthogroups have 26 branches suggesting no divergence at synonymous sites and 36 branches in the tree indicating $dN/dS>>1.0$, a strong signature of natural selection driving amino acid substitutions \citep{Nei1986,Goldman1994}. These results suggest strong adaptation at detox gene families in \Mnerv. We also identify 11 incidences of Hsp70 formation, 21 branches for Opsin genes, 11 mitochondrial-eating protein duplications, 15 branches for von Willebrand proteins, and 5 chitin duplications associated with $dN/dS>1$.  These signatures of rapid amino acid substitution in excess of rates expected under neutrality indicate that new gene formation through gene family expansion makes a significant contribution to adaptation in \Mnerv.  It is difficult to resolve evolutionary events in near time using a single genome. However, we can infer that gene copies where $dS<10^{-6}$ occurred within the last 100 generations  ($10^{-6}/(2* 5\times10^{-9})$) or 2000 years.   

\begin{figure}
\begin{center}
\includegraphics[width=0.4\textwidth]{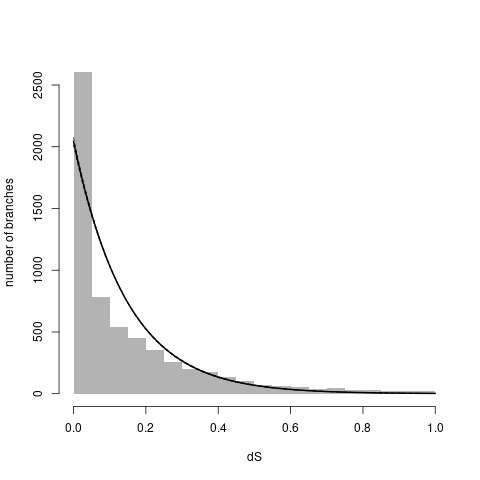}
 \caption{Gene family member ages across the genome of \Mnerv, are fit with a birth-death model.  The distribution of duplicate ages, given by dS, shows larger numbers of younger gene family members, tapering off exponentially as expected for a Poisson process. Model fit suggests a birth rate $\lambda=40,859$ (95\% CI 40,262-41,455) and death rate $\mu=6.8$ (95\% CI 6.7-6.9) per 1.0 dS.  Model fit is independent of histogram binning or visualization.  \label{birthdeath}}
 \end{center}
 \end{figure}

We fit a birth-death model to 6002 first order paralogs for all gene families using dS as a proxy for age according to $\lambda e^{-\mu dS}$ (Table \ref{FamBirthDeath}, Figure \ref{birthdeath}).   The distribution of dS values follows a pattern of exponential decay as expected under birth-death processes (Figure \ref{birthdeath}).  Using maximum likelihood, we fit a death rate, $\mu=6.8\pm$ (95\% CI 6.7-6.9) and a birth rate $ \lambda=40,859$ genes per 1.0 dS (95\% CI 40,262-41,455).  Assuming gene copies appear in clock-like fashion and each accumulate point mutations at a rate of $5\times 10^{-9}$ \citep{scallop}, this birth rate scales to a mutation rate of $1.16 \times 10^{-8}$ duplications per gene per generation.  Such mutation rates are equivalent to copying each gene in the genome once on average every 11 million generations. The half-life of a newly formed gene family member is 0.1 dS, corresponding to 1 million generations or roughly 200 million years.  We note that this is far longer than we found for duplicate gene pairs in our previous work in deletion-biased \Dros {} \citep{Rogers2009}.  The distribution of dS also does not show evidence of long term preservation suggesting an escape from birth-death processes (Figure \ref{birthdeath}).  

\begin{figure}
\begin{center}
\includegraphics[width=\textwidth]{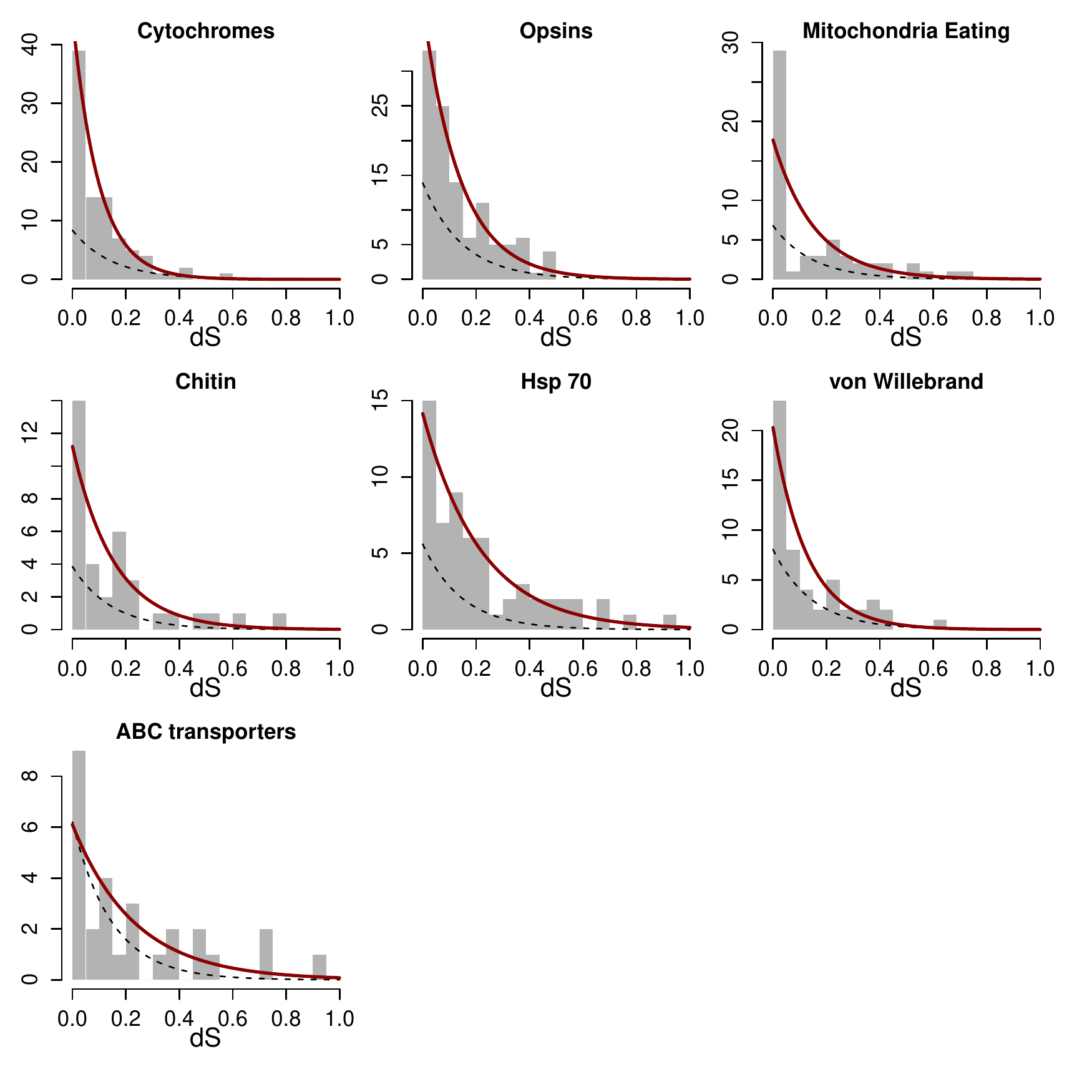}
 \caption{Gene family member ages for high copy number gene families in \Mnerv. We fitted a birth-death models (red lines) to identify families with rapid gain and loss compared with genomic background (dashed lines).  Cytochrome P450 genes show high rates of birth and death compared with background levels with over 7.6-fold increased birth rates ($P<0.0001$). Other gene families have 3.5-4.0 fold higher birth rates, except ABC transporters.  ABC transporters and Hsp70 genes show reduced death rates, with a half-life up to 50\% longer than background. \label{cypbirthdeath}}
 \end{center}
 \end{figure}

We identified 88 \emph{cytochrome P450} genes that appear to be members of large gene families in a first-order Fuzzy Reciprocal Best Hit BLAST \citep{TripleHan}. We estimated dN and dS in PAML and fit birth-death models to these large gene families (Figure \ref{cypbirthdeath}).  For \emph{cytochrome P450} genes, $\lambda=898$, $\mu=10.2$ (Table \ref{FamBirthDeath}).  Scaling birth rates to be proportional to the number of genes examined (143 cytochrome annotations vs 49,149 total genes), this is a  7.6-fold higher rate of new gene formation and a 1.5-fold higher rate of gene family contraction in cytochromes than we would expect based on the genome wide background (Table \ref{FamBirthDeath}, Figure \ref{cypbirthdeath}), well beyond the 95\% CI ($P< 0.0001$).   Among other large gene families,  von Willebrand factors, chitin metabolism genes, Hsp70 genes, mitochondrial eating proteins, and opsins show birth rates 3.5-4.0 times higher than background levels (Table \ref{FamBirthDeath}, Figure \ref{cypbirthdeath}).  ABC transporters and Hsp70 genes appear to be affected by lower death rates, with a half-life up to 50\% longer than background (Table \ref{FamBirthDeath}, Figure \ref{cypbirthdeath}). 

\subsubsection*{Burst of Recent Transposable Element Activity}
TEs are selfish genetic elements that copy themselves often at the expense of their host organisms.  Transposable elements are key drivers of gene family expansion as they can facilitate ectopic recombination, create new retrogenes, and duplicate adjacent gene sequences \citep{oliver2009,Feschotte2008}. The way that these modes of mutation interact to reshape genomes during adaptation to environmental challenges is key to understanding how organisms evolve in nature. 

\begin{figure}
\begin{center}
\vspace{-.35in}
\includegraphics[scale=0.17]{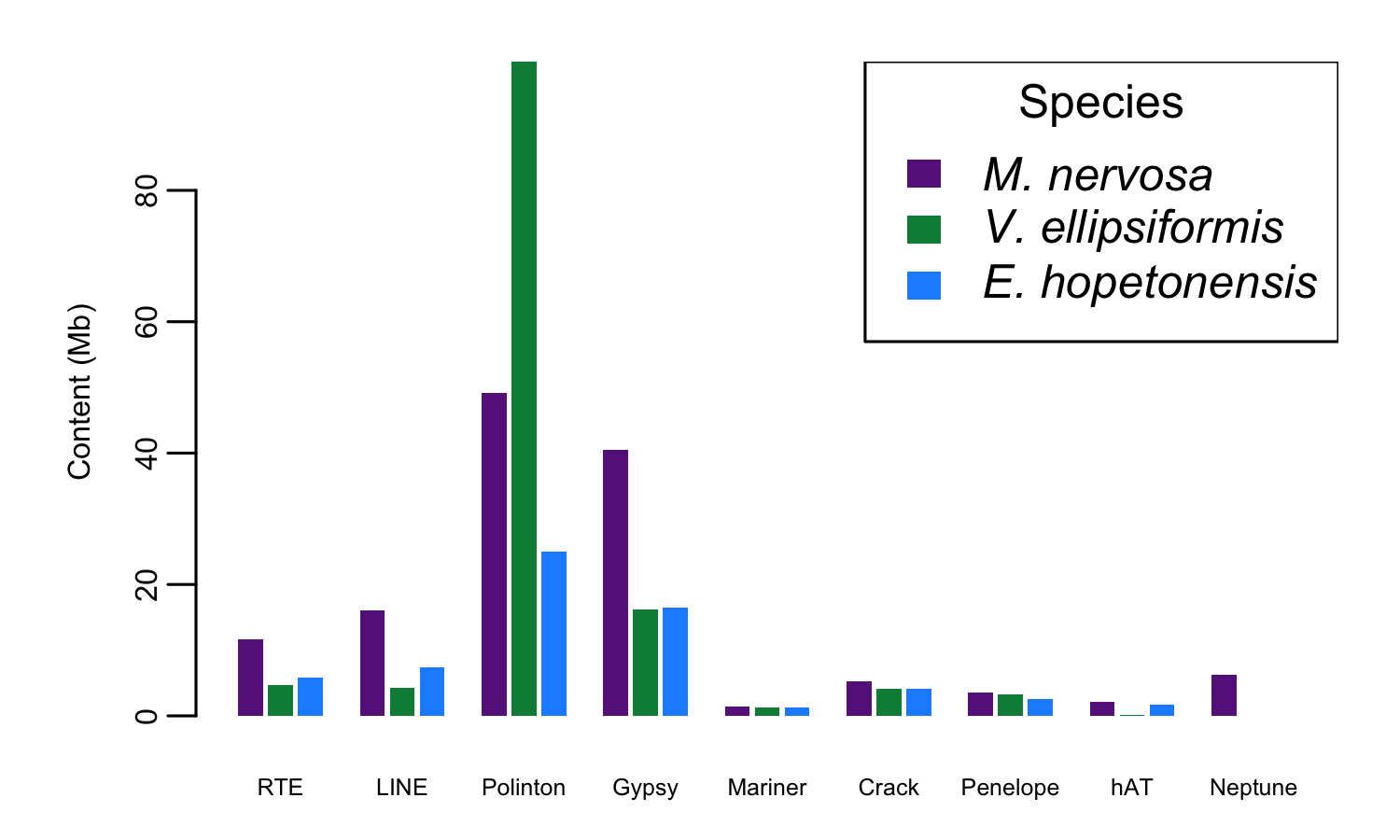}
\end{center}    
\vspace{-.3in}
\caption{TE content for recently proliferating repeat families found in three species of \Unio {} using RepDeNovo, an assembly-free repeat identifier.  \MnervFull {} has the highest aggregate TE content across all families with 382 Mb of new genome content.  \emph{Gypsy} elements and \emph{Polinton} elements each explain 21\% of the expansion in genome content.  A burst of \emph{Polinton} amplification is observed in \Velli. \label{TEs}}
\vspace{-0.1in}
\end{figure}

These high copy number repeats can be identified in next generation sequence data independently from genome assemblies or in fully assembled reference genomes.  We used RepDeNovo \citep{Chu2016} to identify young, newly propagated transposable elements in three species where genomic sequence data are currently available \Mnerv, \Velli, and \Ehope.   We observe evidence of a burst of TE activity in \Mnerv {}, resulting in an expansion of genome content.    We observe roughly two fold greater TE content in \Mnerv {} compared with the other species (Figure \ref{TEs}).  Because these methods do not rely on genome assembly quality, results from RepDeNovo should be directly comparable across species.  These TEs account for a 21\% (382 Mb) increase in genome size compared with previously sequenced bivalve \Velli.  

A total of 21\% of the increase in TE content in \Mnerv {} compared with \Ehope {} is explained by expansion of \emph{Polinton}-like DNA transposons, and 21\% is explained by expansion of \emph{Gypsy} LTR retroelements (Figure \ref{TEs}).  Remaining classes each explain 0-5\% of TE expansion.  Repeat content in \Velli {} roughly resembles that of \Ehope {} with the exception of the 4-fold increase in \emph{Polinton} DNA transposons and a reduced number of \emph{hAT} repeats (Figure \ref{TEs}). Notably, the \emph{Neptune} family of \emph{Penelope}-like elements is identified in \Mnerv {} but absent in \Ehope {} and \Velli, explaining 5\% of the expansion in TE content (Figure \ref{TEs}). Roughly 45\% of the repetitive content in each genome could not be matched in a tBLASTx, suggesting some repeat families are highly diverged from annotated repeats in RepBase \citep{RepBase}.  The difference in genome size for \Velli {} and \Mnerv {} is primarily explained by the proliferation of 382 Mb of repeats.   \Mnerv {} and \Ehope {} are from 42X and 14X coverage data, respectively.  Downsampling genomic data to 14X, total TE content increases marginally to 465 Mb.  Hence, we do not believe the observed differences in TEs between species can be explained by sequence depth but rather reflect biological differences between \Mnerv {} and the two other bivalve species. 

\begin{figure}
\begin{center}
\vspace{-.35in}
\includegraphics[scale=0.5]{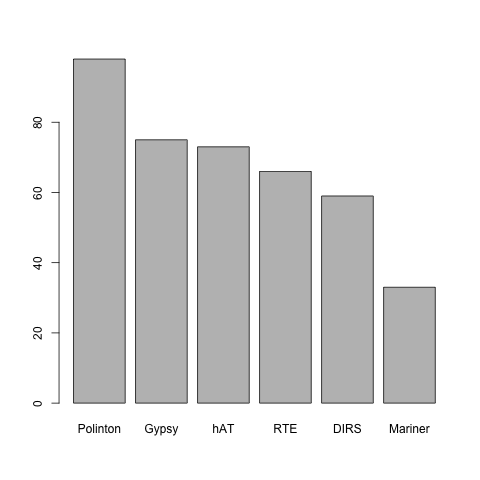}
\end{center}    
\vspace{-.3in}
\caption{Number of repeat sequences identified in the \Mnerv {} reference assembly using RepeatScout for the 6 most common TE families in \Mnerv.  \emph{Polinton} and \emph{Gypsy} elements harbor the greatest number of independent repeats, consistent with results showing ongoing proliferation in assembly-free methods.  Other prevalent elements include \emph{hAT}, \emph{RTE}, \emph{DIRS} and \emph{Mariner} elements.  \emph{LINE} elements showed little representation among repeats.\label{TEs2}}
\vspace{-0.1in}
\end{figure}

RepeatScout identifies older repetitive elements with sufficient divergence to be assembled correctly into genome scaffolds \citep{RepeatScout}.     We identify 3686 repeats in \Mnerv, 1827 of which are greater than 200bp long.  We are able to  classify 573 of these into families using a tBLASTx comparison against RepBase at an E-value of $10^{-10}$.  We find 98 \emph{Polinton} elements, 75 \emph{Gypsy} elements, 73 \emph{hATs}, 66 \emph{RTE}s, 59 \emph{DIRS}, , and 33 \emph{Mariner} elements (Figure \ref{TEs2}).   Other families represented by multiple elements include \emph{Crack}, \emph{Kolobok}, \emph{Nimb}, \emph{Chapaev}, \emph{piggyBac} and \emph{Penelope}.  These elements encompass 275 Mb of sequence.   Diversity for these older groups of TEs compared with results of RepDeNovo would suggest that \emph{hAT}, \emph{Mariner}, and \emph{RTE} elements may have been more active in the ancient past but silent in more recent generations. Meanwhile \emph{Gypsy}, and \emph{Polinton} elements have maintained steady rates of activity. RepeatScout depends directly on the quality of genome assembly, and may not be appropriate for cross-species comparisons unless assembly quality is equivalent. 

We identify only 7 \emph{LINE} elements, with limited diversity to L1 and L2 retroelements, raising questions about when these selfish TEs may have appeared in the phylogeny of \Unio.  Interproscan analysis reveals hundreds of open reading frames for TE-associated protein domains, many with support in the transcriptome (Table \ref{TETerms}). These ORFs have been removed from gene annotations described above.   In total, the results of RepDeNovo and RepeatScout suggest that \Mnerv {} harbors a total of 575 Mb of transposons, roughly 25\% of genome content. 

\subsubsection*{Effective Population Size}
The tempo and outcomes of evolution via selection and drift depend directly on the effective population size.  Coalescent theory suggests that heterozygosity, the number of differences between chromosomes in diploid individuals per site, will be distributed according to $H=\theta=4N_e\mu$. Here, $N_e$ is the effective population size necessary to produce genetic diversity in an idealized model and $\mu$ is the mutation rate.  Modern population genetic theory is based on these parameters and the ways that they shape the response to forces of drift and selection in natural populations.  We can use theoretical models to explore the limits of current genetic diversity and the likelihood of adaptation from SNPs and gene duplications.

\begin{figure}
\begin{center}
\includegraphics[width=0.4\textwidth]{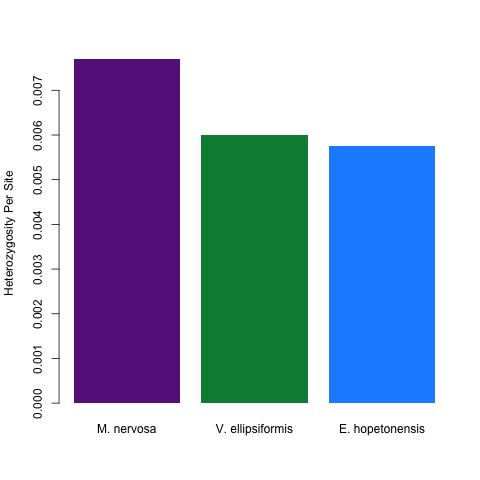}
\caption{Heterozygosity in \Mnerv,  \Velli, and \Ehope.  \MnervFull {} shows the highest heterozygosity, consistent with its expansive range across the Southeastern United States and Northern Mississippi River.  \label{het}}
\end{center}

\end{figure}

We have used heterozygosity in reference specimens for three species of bivalves to estimate $\theta$. We observe high heterozygosity suggesting $\theta_{M.nervosa}=0.00777$, $\theta_{V.ellipsiformis}=0.0060$, and $\theta_{E.hopetonensis}=0.00576$ (Figure \ref{het}).  Results from \Mnerv {} match well compared to within-population estimates for a small number of loci \citep{Pfeiffer2018}.  Our results suggest that genetic diversity in these three species of bivalves is still high, even in a species that has a limited geographic range and more specialized fish host influencing dispersal \citep{Johnson2012}.  \Ehope {} is an endemic to Georgia and South Carolina with a limited range but with census sizes estimated to be over $10^6$ individuals.  It is not surprising that it would have such lower $N_e$ than \Mnerv {} and \Velli, which are widespread throughout the Eastern United States, if migration offers connectivity across drainages.  These observed levels of heterozygosity lead to estimates for $N_e$ of 388,500 individuals in \Mnerv. In \Velli {} $N_e$ is estimated at 300,000 individuals and in \Ehope {} 288,000 individuals.  These estimates are substantially higher than estimates that have been observed in humans of roughly 15,000 individuals \citep{PSMC} but slightly lower than $10^6-10^8$ is seen in \emph{Drosophila} \citep{Karasov2010} or $10^7$ or greater in unicellular eukaryotes \citep{LynchBook}.  

We estimate coalescent times based on $2N_e$ generations of 777,000 generations for \Mnerv, 600,000 generations for \Velli, and 576,000 generations for \Ehope {} (Table \ref{Pgen}) with $t_{MRCA}$ scaling by 2 times higher for $4N_e$ generations  \citep{Wakeley2009}.  This evolutionary system is unusual in comparison with life history of organisms commonly used for population genomics like yeast and \Dros {} (large $N_e$, short generation time) or humans (small $N_e$, long generation time) \citep{LynchBook}.   It is rarer to find systems where populations have housed millions of individuals with long generation times.  The observed high $N_e$ in a long lived, late maturing species is expected to influence the tempo of evolution.  Given a mutation rate of $5\times10^{-9}$ \citep{scallop} and a modest estimate of 20 years per generation, the coalescent times under neutrality for such diversity would span up to 15 million years and total population diversity by 30 million years based on $t_{MRCA}$.  

\subsubsection*{Gene tree phylogenies for \Unio}
\begin{figure}
\begin{center}
\includegraphics[width=1.0\textwidth]{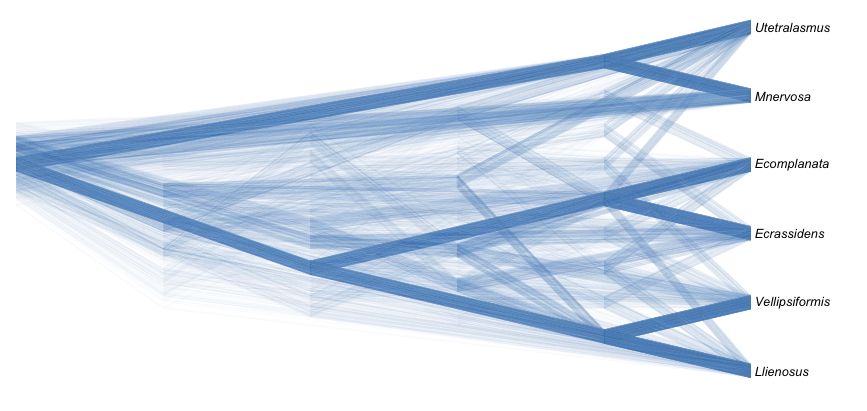}
\caption{ Gene tree topology for 6 species of \Unio.  The consensus tree matches previous mtDNA phylogenies from Campbell et al. 2005.   The consensus tree is supported by 47.6\% of gene trees.  Terminal bifurcations suggesting 2-way species splits are supported by 66.2-73.3\% of gene trees.  Final normalized quartet score is 0.7605491438485329 from Astral III, where ~76.06\% of all gene quartets match species tree. Gene trees that incongruence driven by incomplete lineage sorting, introgression, or molecular convergence is common for 33-50\% of gene trees.  Branch lengths are unscaled for purposes of visualization.  \label{DensiTree}}
\end{center}
\end{figure}

The observed value of $\theta=0.0077$ in \Mnerv {} raises the prospect of long coalescent timescales that may in fact be expected to transcend events of species separation.  As a first step to examine relationships among \Unio {} and capture information about potential incomplete lineage sorting we produced gene-tree phylogenies for 6 species where transcriptome sequences were available and of sufficient quality to facilitate phylogenetics: \Mnerv, \Utet, \Ecomp, \Ecras, \Llien, \Velli.    Using \emph{de novo} assembled transcriptomes we identified 4475 genes where 1:1 orthologs were present across all 6 species with high quality sequence alignments.  We placed the root of the tree at the divergence of Quadrulini, according to prior mtDNA based phylogenies from \cite{Campbell2005}.  The consensus across all gene trees is consistent with the phylogeny represented in this previous work on mtDNA, indicating that the maternal haplotype tree reflects the species tree for these six species.  A total of 2131 out of 4475 gene trees or 47.62\%, are fully consistent with the consensus species tree  (Table \ref{GeneTrees}, Figure \ref{DensiTree}).  For any bifurcation at the terminal branches 66.2\%-73.3\% of gene trees support the three terminal species splits  (Table \ref{GeneTrees}, Figure \ref{DensiTree}).  Final normalized quartet score is 0.7605 from Astral III, where ~76.05\% of all gene quartets match species tree.  Unscaled branch lengths are plotted for purposes of visualization, but scaled branch lengths are available in Supporting Information. We were unable to place a 7th species \Aplic {} in any consensus tree.  It is unclear whether this species if affected by complex history driving gene tree incongruence or bioinformatic and data quality artifacts (see Supporting Information).

Some 52.4\% of gene trees are discordant with the consensus tree in some way.  Gene tree discordance can be caused by incomplete lineage sorting (ILS) when coalescent times are so distant in the past that they predate speciation events.  These processes occur when random assortment of ancestral polymorphism results in inheritance patterns that contradict the true species tree.  Gene tree discordance can also occur when interbreeding among species allows alleles to introgress across species boundaries.  Discordant gene trees from ILS should be symmetric with respect to species, as the assortment of polymorphism into clades occurs at random.   Introgression, however, typically leaves a signature of asymmetric agreement with one phylogenetic clustering over the other.   We observe two species clusters that show such asymmetry, one in the placement of \Utet, the other in the clustering of \emph{Elliptio} with \Velli.  Speciation events appear on the order of 5000-1000 generations or roughly 100,000-200,000 years, less than the coalescent timescale.  Short branch lengths among species have been previously noted as one impediment to clear phylogenies \citep{Pfeiffer2019,Araujo2018} and may increase the chances of gene tree, mtDNA, and species tree discordance \citep{Degnan2009,moore1995,toews2012}. 


\subsubsection*{Adaptive potential in \Mnerv}
We can estimate the time necessary for a new allele to appear and establish a selective sweep in a population, $T_e$.  We observe waiting times for strong selective sweeps ($s=0.01$) on new mutations ranging from 12,870 generations in \Mnerv {} and 56,180 generation in \Ehope.  Using an estimated generation time of 20 years, these numbers correspond to 257,000 years in \Mnerv {} and 347,000 years in \Ehope {} (Table \ref{Pgen}), exceptionally long wait times to respond to fluctuating environments.   Under stronger selection coefficients of $s=0.10$, we would expect wait times of 25,700 years in \Mnerv {} and 34,700 years in \Ehope {} (Table \ref{Pgenhigh}). Hence, we would suggest that any response to modern anthropogenic selective pressures in \Unio {} will necessarily proceed from standing variation.  

The probability of adaptation from standing variation at single nucleotide sites is given by $P_{sgv} = 1 - e^{-\theta*ln(1+2N_e s)}$ \citep{Hermisson2005}.  These equations account for the probability that a given mutation is available within a population and that it exists at high enough frequency to establish a deterministic selective sweep.  Given estimates of $N_e$ and $\theta$ observed in these three cursory reference genomes, we estimate probabilities of fixation from standing variation under a selection coefficient of 0.01 of 4.8\% for \Ehope, 5.1\% for \Velli, and 6.7\% for \Mnerv.  \MnervFull, with its high genetic diversity will hold a 33\% advantage in the ability to respond to sudden shifts in selective pressures compared with endemic species like \Ehope.  Substituting the mutation rate observed for new gene formation  estimated from birth-death models ($1.16 \times 10^{-8}$ duplications per gene per generation) and identical parameters of $N_e$ and $s$, we would estimate $P_{sgv}$ for gene family expansion to be 18.5\%.  Hence, gene family expansion is expected to be an important substrate of novelty in \Mnerv.  

\subsubsection*{Discussion}
 \subsubsection*{Gene family expansion and natural selection}
 Gene family evolution is an important source of novel genetic material that produces innovation in the face of shifting selective pressures \citep{conant2008,Ohno1970,Hahn2009}.   New genes arise through duplication, then accumulate mutations that modify original functions \citep{conant2008}.  They may also modify expression patterns, especially when regulatory elements are shuffled in the face of partial duplication \citep{Rogers2017Exp}.  In the face of rapidly changing selective pressures, these substrates of genetic novelty are expected to be all the more essential as organisms require adaptive innovation.  In \Mnerv, we estimate rates of gene duplication equivalent to $1.16 \times 10^{-8}$ per gene per generation.  Such a high mutation rate combined with the large population sizes estimated for \Mnerv {} suggest a high probability of adaptation from standing genetic variation of $P_{sgv}=0.185$.  This estimate is 2.76-fold greater than $P_{sgv}$ for single nucleotide changes (Table \ref{Pgen}).  Hence, the data suggest a greater likelihood of adaptation from standing genetic variation for any given gene duplication compared with any given SNP.  This theoretical rate suggests that adaptation through gene family amplification may offer a rich substrate to respond to environmental changes, especially when new functions are needed.  
 
The \Mnerv {} genome houses an unusually high rate of copy number expansion in detox genes compared with genome wide background rates of duplication, with a birth rate 7.6 times higher.  Hence, we conclude that cytochromes are operating under enhanced birth-death processes with rapid expansion (and possibly contraction) in \Mnerv {} compared with genome wide expectations. This enhanced birth rate for new gene formation is likely to be responsible for the observed excess of \emph{cytochrome P450} genes compared with many other annotated animals.   Humans have only 27 cytochrome P450 genes while mice have 72 \citep{Nelson2004}, half the number observed in \Mnerv.  Moreover, 26 \emph{cytochrome P450} genes appear with no divergence at synonymous sites, suggesting recent amplification within the past 100 generations (roughly 2000 years) though timescales are difficult to resolve in the near term.  Determining more precisely whether these changes preceded the modern industrial era and damming of the river in 1918 would require data from more individuals, potentially from historic specimens.  Regardless it is clear that detox gene amplification in \Mnerv {} has expanded gene content and aided adaptation in recent evolutionary time, whether they appeared among standing variation before or after the anthropogenic era. 
  
  Alongside these newly formed copies, many of the remaining copies appear to be older, suggesting pre-adaptation or genetic priming for highly toxic environments. The cytochrome gene family houses 36 duplication events with high $dN/dS>1.0$.  Such signatures indicate rapid amino acid substitutions in excess of expectations under neutrality, an indication of strong selective pressures and adaptive processes driving amino acid substitutions \citep{Nei1986,Goldman1994}.  The Tennessee River has been historically subject to high pesticide loads from TVA's mosquito control programs and from cotton farming \citep{woodside2004}.   Previous annotation of \Ecomp {} has found amplification of 12 \emph{cytochrome P450} genes by annotation via homology with \Cgig, whose long divergence time is likely to limit the number of copies that can be identified \citep{Cornman2014}. 
  
  Such expansion occurs in addition to observed signatures of amino acid substitutions driven by selection.  In aggregate, these results of gene family evolution point to widespread adaptation through gene family expansion, especially in detox genes.  Toxicity assays in \Unio, including \Mnerv {} have identified high tolerance for multiple pesticides and herbicides \citep{milam2005,loganathan1998}, consistent with high rates of exposure expected in filter feeders that process large water volumes \citep{bril2017}.  Toxicity levels in bivalve tissues have been noted in excess of 0.2 ppm \citep{bedford1968}, similar to concentrations that can be used as bait to kill insects \citep{Schmidt2017}.  
 
 Large gene families with new copies are also identified among anticoagulation genes consistent with blood feeding in the parasitic larval stage, though these may be easier to identify in \Mnerv {} with transcriptomes for glochidia.  We note expanded families of chitin metabolism genes, mitochondria-eating proteins, Opsins, and heat shock proteins.   \Unio {} are known for dual uniparental inheritance of mitochondria, where males inherit germline mitochondria from the father and females inherit them from the mother.  The excess of mitochondria-eating genes and cytochromes that are part of electron transport chain seems intriguing in light of this unique biology.  It is possible that these genes may play a role in regulating mitochondria, especially given the known heteroplasmy in male \Unio {} and processes of DUI \citep{Breton2007, Breton2011, Liu1996, Wen2017}.  Whether these genes are directly involved in DUI or rapidly evolving in response to DUI may offer an area of new exploration in \Unio {} genetics.  These large gene families have 3.5-4.0 fold higher birth rates than genome background rates suggesting rapid gene gains.   
  
 By identifying new genes and searching for genetic signatures of rapid evolutionary events, we can facilitate reverse ecological genomics and discover evolutionary processes that are important to the organism without pre-existing human biases.  While we placed no prior constraints on functional categories, we note many large gene families associated with \Unio {} biology.    Globally, these results suggest that new gene formation has had a substantial impact on the genome of \Mnerv. It  has experienced widespread creation of new genes with strong signatures of adaptive evolution and recent evolutionary changes.  The ways that these mutations contribute to adaptive changes in the near term are well worth further study to determine how they contribute to \Unio {} evolution as well as their importance in evolutionary theory.

 \subsubsection*{Transposable elements and genome remodeling}
 
 TE mobilization can lead to rapid and dynamic changes in genomes \citep{oliver2009,Feschotte2008}. Selfish elements can copy, modify, or break gene sequences as well as induce regulatory changes in neighboring genes \citep{oliver2009,Feschotte2008}.  These mutations are most often detrimental \citep{LynchBook}.  They can however incidentally create new genetic variation within genomes that shapes evolutionary outcomes.  TEs are known to activate in times of organism stress.   It has been proposed that TE activation can be a source of gene expression remodeling to modify suites of genes in few mutational steps \citep{oliver2009,laudencia2012,Lynch2011}.  These TEs are related to the  high rates of gene family expansion observed in \Mnerv, as mobile elements are expected to facilitate new gene copies through gene capture, retrogene formation, and ectopic recombination.   The fact that both TEs and gene families exhibit dynamic evolution suggests that these mutations have an exceptional impact on genome remodeling and evolutionary change.   
 
 We observe a recent burst of TE mobilization for \emph{Polinton} and \emph{Gypsy} elements in \Mnerv {} that resulted in a 382Mb expansion in genome size.  Comparisons with older transposons in the genome assembly suggest that these families have had ongoing proliferation throughout the history of \Mnerv.  In contrast, \emph{hAT}, \emph{Mariner}, and \emph{RTE} elements appear to have been active in the more ancient past but largely quiescent in the recent past.  The causes and impacts of these changing TE dynamics deserve further exploration in future cross-species comparisons as well as searches for variation within populations.  
 
 It remains unclear to what extent the recent TE mobilization in \Mnerv {} may have produced beneficial changes, neutral variation, or widespread damage to genomes, a prospect that might be addressed in future population genetic work.  With modern population size reduction \citep{Ahlstedt1992}, the nearly-neutral coefficient will have changed by more than one order of magnitude.  It is possible that this shift in nearly-neutral impacts \citep{Ohta1973} has allowed some TEs to proliferate unchecked \citep{LynchBook}, even while absolute population sizes have remained very large.  It is equally possible that strong selection has favored some insertions or TE mediated genome remodeling as a means to respond to environmental challenges.  The observed adaptive signatures and rapid evolution of gene families may point to an interaction between TEs and adaptation through duplication, a prospect that deserves future exploration. 
 
 Regardless of whether this TE expansion has produced beneficial or detrimental changes, it is clear that recent proliferation of TEs has altered genome structure and expanded genome content in \Mnerv {} compared with its relatives.  It is possible that TEs have proliferated in the face of environmental stressors, declining populations, or even due to invasion via horizontal transfer from other organisms \citep{Schaack2010}.   However, it is also possible that some of these new TE insertions reflect adaptive changes as selfish genetic elements remodel genomes and incidentally create new genes or new expression profiles across the \Mnerv {} genome.  Sorting out the adaptive TE insertions from detrimental TE insertions will require full population genomic datasets, a prospect that will become tractable with this new reference genome.

\subsubsection*{Evolutionary relationships in \Unio}
 Phenotypic studies in freshwater bivalves have noted rampant convergence of forms across lineages \citep{Smith2020,Pfeiffer2016,Inoue2013,Johnson2018,williams2008}, and phylogenetic analysis has been challenging in the face of variation rather than adherence to strict phenotypes \citep{Campbell2005}.  Our preliminary analysis of gene trees based on transcriptome assemblies for 6 species are consistent with these previous challenges of phylogenetic placement based on both phenotypes and genetic analysis of few loci.  The population genetic parameters for three species of \Unio {} where whole genome sequences are available also imply that the coalescent time scale may be greater than many species splits.  Under these circumstances we may expect rampant trans-species polymorphism and widespread convergence from adaptation via shared ancestral variation.  
 
 In transcriptome data, we observe a clear consensus tree among all gene trees.  However, over 50\% of gene trees differ from the consensus in some way.   Many discordant gene trees are symmetric with respect to species placement, a feature of ILS.  However we observe asymmetric relationships between gene trees and species trees at some nodes, raising the question of how much interbreeding may have occurred across species recently or in the past.  Curiously, a 7th species, \Aplic, does not have a clear placement in the phylogeny.  Phylogenetic discordance has been noted in the face of short branch lengths in \Unio {} in broad phylogenetic surveys using bait pulldown sequencing \citep{Pfeiffer2019}. 
 
 Focusing on gene trees for 6 focal species, we are able to offer finer resolution and a more focused account of incomplete lineage sorting among closely related species.   The observed discordance even among these few species is consistent with predictions based on population genetic diversity measures and long coalescent times.  Full resolution of how complex histories of ILS, introgression, and convergence influence \Unio {} remains an open question that will require the field to move beyond transcriptome level analysis.    With future work developing high quality genome assembly and annotation for more species across the diversity of \Unio, it may be possible to place quantitative bounds on how these evolutionary forces have shaped the genetic underpinnings of phenotypic diversity in \Unio. 

\subsubsection*{Tempo of Evolution}

The tempo of evolution depends directly on effective population sizes and generation times.  We have observed high heterozygosity in \Unio, with $\theta=0.0077$ per site in \Mnerv.   Under these parameters, we expect neutrally evolving variation will experience long coalescent times of 777,00 generations, up to 15 million years assuming constant generation times of 20 years.  Population genetic data can be used to extract information concerning population history up until the $t_{MRCA}$ of all individuals in the population, which is expected to be 4N generations \citep{Wakeley2009}, or 30 million years.   This timescale for neutral variation spans ice ages and mass extinctions, encompassing historical environmental changes that have reshaped the history of the earth.   Previous work on a small number of genetic loci has noted high levels of sequence variation within many populations of \Mnerv {} and limited differentiation across geographic areas \citep{Pfeiffer2018}.  

It is suggested that the larval life history as generalist fish parasites may allow migration across multiple habitats, increasing diversity within populations to overcome geographic separation across its range \citep{Pfeiffer2018}, a prospect that is likely to be important for genetic diversity and conservation in largely sessile species \citep{Archambault2018}.  We observe the highest genetic diversity in \Mnerv, and the least in the endemic species \Ehope.  M. nervosa is known to use lots of common fish as hosts \citep{woody1993,williams2008}.  However, not all \Unio {} are broadcast spawners, such as some Elliptios, which cast mucus nets and use fewer fish species for dispersal \citep{Johnson2012}. This difference in host strategy within broadcast spawning mussels probably impacts dispersal. These life history factors likely play an additional role in shaping genetic diversity, especially differentiation across drainages within the species range. 

Populations of bivalves from Northern and Southern Alabama are separated by the Fall Line, where river systems flow from rocky upland provinces to the less consolidated Coastal Plain \citep{williams2008}.  Periodic connections when rivers varied courses across waterways as glaciers worldwide formed and receded allowed for species migration far in the past, potentially influencing genetic diversity.  In light of these numbers, it appears likely that the historical and geological forces that have shaped these species may indeed still be written in their DNA.  This long-standing genetic diversity may offer the chance to use population genetics to explore evolutionary theory in deeper timescales than most standard population genetic models.
 
The genetic diversity of \Mnerv {} and other \Unio {} is shaped by ancestral variation, and is expected to decay gradually over time as populations diminish or become fragmented \citep{Wright1931,Wright1951}.  Still, in the face of modern anthropogenic selective pressures in \Unio, this genetic variation will be essential for adaptation.  Population genetic theory predicts that selective sweeps on new mutations will be slow to appear, requiring tens of thousands of generations or hundreds of millions of years for desired genetic variation to appear \citep{Gillespie1991,Maynard1971}.  For adaptation to occur on nearer timescales, standing variation is expected to offer the substrate for initial adaptive responses.   The probability that a genetic variant exists and is able to establish a deterministic selective sweep in a population $P_{sgv}$, depends directly on $\theta$ \citep{Hermisson2005}.  In these bivalves, \Mnerv {} has the highest level of genetic variation, 35\% higher than endemic \Ehope.   This will result in a 33\% higher chance of adaptation from standing variation, a key advantage in the response to ongoing selective events.   With modern population reductions \citep{Ahlstedt1992}, this ancestral standing variation will be even more essential than these estimates from historic population sizes might reflect.  

Census of bivalves noted smaller classes of \Mnerv {} below the 4 inch commercial harvest limit were noted to be scarce in Wheeler Lake \citep{Ahlstedt1992}, though the authors have noted healthy numbers of juveniles near Pickwick Lake.  It is unlikely that $N_e$ fully reflects the ongoing mussel declines, nor do we expect it reflects equilibrium levels for reduced populations.  It does, however, indicate that the species still house substantial genetic diversity, if populations can be preserved.  We would expect a decay in this diversity over time as population sizes drop according to $H_t=H_0 (1-\frac{1}{2N})^t$ \citep{Gillespie2004}.   Even assuming the minimum age of reproduction at 5 years of age, it has been fewer than 20 generations since the damming of the Tennessee River began in 1918 and even fewer since Pickwick Dam was established in 1938.  Hence, we expect that genome wide $H_0$ will reflect historic rather than modern diversity.  

The extent to which ongoing trace migration across the geographic range, especially incidental or intentional transport with humans, might reseed population diversity remains largely understudied.  Previous work has suggested that diversity in \Mnerv {} is well connected across most reservoirs, potentially benefiting from diversity of fish hosts that can transport offspring \citep{Pfeiffer2018}.  Future work with whole genome sequencing multiple individuals across the geographic range might clarify whether populations remain panmictic or if distinct genetic features may be emerging in specific geographic locations via drift or local adaptation. 

Here, \Unio {} offer the prospect to study how adaptation occurs in long-lived but numerous species that is notably different from many population genetic models.   Yet, the implications of these differences go far beyond abstract concepts in population genetic theory.  Those organisms that house the greatest diversity may indeed have the best chances for adaptive evolution that could help them survive when new genetic variation is needed.  If adaptation from new mutations requires long timescales, standing genetic variation is the major source of adaptive changes in the near term.  If standing variation is sufficient to offer genetic solutions to selective pressures, organisms will be able to readily adapt.  If genetic diversity is unavailable, animals faced with strong selection will be unable to realize adaptive change and may be at greater risk of extinction.  Indeed, standing genetic variation has been shown to be essential as other marine bivalve species adapt to anthropogenic changes \citep{Bitter2019}.  Evaluating the substrate of genetic variation is essential to understand the limits that these species face, especially given widespread extinction of freshwater \Unio.  

\subsubsection*{Genetic resources for freshwater \Unio} 
 
\Unio {} currently face a modern day mass extinction \citep{williams1993, Lopes2018, Haag2014}.  They have faced strong recent selective pressures of pollution, barriers on waterways, invasive species, and loss of fish hosts.  These organisms badly need genetic resources to assess population health, monitor diversity, and identify unique groups across geographic locations.   Moreover, evolutionary analysis is currently limited by a lack of genetic resources.  At present, a single fragmented whole genome sequence exists across all \Unio {} for comparison \citep{Renaut2018}, and the nearest outgroup, \Cgig {} is 495 million \citep{TimeTree} to 200 million years \citep{bolotov2017}.  

Previous work has used limited mtDNA sequencing, targeted single genes, transcriptomes, or few microsatellite markers \citep{Campbell2005, Pfeiffer2019,Cornman2014,Gonzalez2015,Wang2012}.  In the absence of whole genome sequencing, bait-based pulldowns of SNPs combined with sequencing have been useful in reconstructing phylogenies \citep{Pfeiffer2019} (though these may be subject to some measure of ascertainment bias).   Each of these approaches offer useful genetic data in cost-effective schemes.   Future whole genome sequencing will offer a more complete description of genomic changes, especially with respect to duplicate gene formation and TE proliferation across \Unio.   The reference genome for \Mnerv {} represents a step toward evolutionary genomics of gene families and TEs in a high-risk clade, offering a portrait of dynamically changing genomes that have responded to strong selection in the past.  This work represents a first step to understanding of genetic responses in a robust species within a larger clade of vulnerable species.  

In this work, we have identified high copy number for key gene families in \Unio {} responsible for detox, mitochondria maintenance, heat shock, and blood feeding. Transcriptome based analysis suggests that comparable gene family expansion is likely to be common among \Unio.  As these functions are directly related to the known life history and environment of the organism, future studies of duplicate genes may be essential to understanding survival and reproduction across the phylogeny.   Genome annotation often masks highly repetitive sequences prior to annotation to reduce the likelihood of identifying TEs as false gene sequences.  This standard practice of masking repetitive sequences prior to annotation would miss many members of these important gene families relevant to the unique biology of freshwater \Unio.  Instead a more focused, precise study of transposable elements with post-annotation filtering is able to sort out this adaptive variation, and offer a more complete description of evolutionary change in \Mnerv, an approach that should be noted in future genome annotation efforts.  
 
Freshwater ecosystems have recently been threatened by widespread upheaval and human encroachment \citep{Lake2000}.   In the modern industrial era, waterways have been dammed throughout the United States limiting dispersal \citep{Ortmann1924}.  Alongside these changes to rivers, waters are compromised with pesticides, industrial pollution, and agricultural runoff \citep{woodside2004}. One unintended consequence has been limited dispersal of aquatic life through waterways, leading to a marked reduction in species diversity and prevalence \citep{Ortmann1924}. TVA has monitored unionids in the Tennessee River for decades as an indicator of water quality and to track impacts of commercial mussel capture that was common until the 1990s, providing a rich historical record \citep{Cahn1936, Isom1973}.  Improved environmental regulations and targeted conservation efforts have encouraged recovery, but the historic mark of these environmental challenges is not fully known.    The addition of whole genome sequence analysis to this rich historical record will enhance conservation and population monitoring in this focal location that is widely recognized for its high diversity of \Unio.
 
These genome sequences can further help target conservation efforts where they are needed most.   \Unio {} with lowest sequence diversity can be marked as first priority for conservation.   Monitoring whole genome genetic variation can help identify a baseline so that genetic diversity is preserved as wildlife management efforts survey freshwater \Unio {} populations.  In the future, even low-level whole genome sequencing to identify nuclear genetic variation is likely to facilitate phylogenetic and evolutionary analysis in \Unio {} and offer a more complete account of genetic diversity in a clade where so many species are threatened.  \Mnerv {} populations have remained hardy so far in the face of many ecological challenges.  They offer a benchmark that can be used for comparison in species whose populations are dwindling. For low diversity species, adaptation to newly arising challenges will be increasingly difficult with a lesser substrate for adaptation.  In these cases, early intervention may be warranted so that their modest diversity remains preserved.  Additionally, when evaluating which efforts are most likely to succeed, species that retain moderate to high diversity may have the greatest chances of surviving the modern anthropogenic challenges.  Similar genetic studies in the future will be increasingly essential for evolutionary theory and conservation in \Unio.  

 
\subsubsection*{Data availability}
\VelliFull {} transcriptome data are available at SRR6279374 and genomic sequence data were taken from SRR6689532 and SRR6689533. \MnervFull {} genomic sequence data are available in PRJNA646917 and transcriptome data are available at SRA PRJNA646778.  \EcrasFull {} transcriptomes are deposited under SRA PRJNA647326.  \EhopeFull {} genomic reads are deposited at SRA PRJNA647325.   Full annotations, \emph{de novo} assembled transcriptomes, and interproscan results are available in Supporting Information, currently on (\href{https://www.dropbox.com/sh/4wqbc7mks1rzajl/AADqDKN6jAWvozMr7Q3cbDiva?dl=0}{Dropbox}).  

\subsubsection*{Acknowledgements}
We thank Rob Reid and James Baldwin-Brown for helpful suggestions about genome assembly software and annotation. Jason Wisniewski provided \Ehope {} specimens, and Robert Bringolf provided \Ecras.  Jon Halter, Michael Moseley, and Chad DeWitt provided help with software installation and functionality. We thank the Duke University School of Medicine for the use of the Sequencing and Genomic Technologies Shared Resource, which provided Illumina sequencing services for \Mnerv {} genomes and transcriptomes. The NC State Sequencing core provided HiC library services.  De Novo Genomics generated Oxford Nanopore sequences for this study.  the University of Colorado Sequencing Core generated sequence data  \Ehope {} and the University of Georgia Genomic Core Facility generated RNA sequencing data for \Ecras.  All analyses were run on the UNCC Copperhead High Performance Computing cluster, supported by UNC Charlotte and the Department of Bioinformatics.  

\subsubsection*{Funding}
This work was supported by startup funding from University of North Carolina, Charlotte Department of  Bioinformatics and Genomics. \Rebekah L. Rogers is funded in part by NIH NIGMS R35 GM133376. The funders had no role in study design, data collection and analysis, decision to publish, or preparation of the manuscript.    The findings and conclusions in this article are those of the author(s) and do not necessarily represent the views of the U.S. Fish and Wildlife Service.

\subsubsection*{Author contributions}
RLR, SP, JET, CCM and JPW designed experiments and analyses \\
JG collected specimens from wild populations \\
RLR, SLG, KB, and CCM performed experiments \\
RLR, SP, JET, and SLG performed analyses \\
RLR, JPW, CCM, JET and JG wrote and edited the manuscript \\

\clearpage

\bibliography{MnervosaRev1Arxiv}

\begin{thebibliography}{}

\bibitem[Adams et~al., 2000]{Adams2000}
Adams, M.~D., Celniker, S.~E., Holt, R.~A., Evans, C.~A., Gocayne, J.~D.,
  Amanatides, P.~G., Scherer, S.~E., Li, P.~W., Hoskins, R.~A., Galle, R.~F.,
  et~al. (2000).
\newblock The genome sequence of \emph{{D}rosophila melanogaster}.
\newblock {\em Science}, 287(5461):2185--2195.

\bibitem[Adrion et~al., 2017]{Adrion2017}
Adrion, J.~R., Song, M.~J., Schrider, D.~R., Hahn, M.~W., and Schaack, S.
  (2017).
\newblock Genome-wide estimates of transposable element insertion and deletion
  rates in \emph{{D}rosophila melanogaster}.
\newblock {\em Genome biology and evolution}, 9(5):1329--1340.

\bibitem[Ahlstedt and McDonough, 1992]{Ahlstedt1992}
Ahlstedt, S.~A. and McDonough, T.~A. (1992).
\newblock Quantitative evaluation of commercial mussel populations in the
  {T}ennessee river portion of {W}heeler {R}eservoir, {A}labama.
\newblock In {\em Conservation and management of freshwater mussels.
  Proceedings of a UMRCC symposium}, pages 12--14.

\bibitem[Aminetzach et~al., 2005]{aminetzach2005}
Aminetzach, Y.~T., Macpherson, J.~M., and Petrov, D.~A. (2005).
\newblock Pesticide resistance via transposition-mediated adaptive gene
  truncation in \emph{{D}rosophila}.
\newblock {\em Science}, 309(5735):764--767.

\bibitem[Araujo et~al., 2018]{Araujo2018}
Araujo, R., Buckley, D., Nagel, K.-O., Garc{\'\i}a-Jim{\'e}nez, R., and
  Machordom, A. (2018).
\newblock Species boundaries, geographic distribution and evolutionary history
  of the western palaearctic freshwater mussels {U}nio ({B}ivalvia:
  {U}nionidae).
\newblock {\em Zoological Journal of the Linnean Society}, 182(2):275--299.

\bibitem[Archambault et~al., 2018]{Archambault2018}
Archambault, J.~M., Cope, W.~G., and Kwak, T.~J. (2018).
\newblock Chasing a changing climate: Reproductive and dispersal traits predict
  how sessile species respond to global warming.
\newblock {\em Diversity and Distributions}, 24(7):880--891.

\bibitem[Bao et~al., 2015]{RepBase}
Bao, W., Kojima, K.~K., and Kohany, O. (2015).
\newblock {R}epbase {U}pdate, a database of repetitive elements in eukaryotic
  genomes.
\newblock {\em Mobile {DNA}}, 6(1):11.

\bibitem[Barnhart et~al., 2008]{Barnhart2008}
Barnhart, M.~C., Haag, W.~R., and Roston, W.~N. (2008).
\newblock Adaptations to host infection and larval parasitism in unionoida.
\newblock {\em Journal of the {N}orth {A}merican {B}enthological {S}ociety},
  27(2):370--394.

\bibitem[Bedford et~al., 1968]{bedford1968}
Bedford, J., Roelofs, E., and Zabik, M. (1968).
\newblock The freshwater mussel as a biological monitor of pesticide
  concentrations in a lotic environment 1.
\newblock {\em Limnology and Oceanography}, 13(1):118--126.

\bibitem[Bennett et~al., 2004]{Bennett2004}
Bennett, E.~A., Coleman, L.~E., Tsui, C., Pittard, W.~S., and Devine, S.~E.
  (2004).
\newblock Natural genetic variation caused by transposable elements in humans.
\newblock {\em Genetics}, 168(2):933--951.

\bibitem[Bitter et~al., 2019]{Bitter2019}
Bitter, M., Kapsenberg, L., Gattuso, J.-P., and Pfister, C. (2019).
\newblock Standing genetic variation fuels rapid adaptation to ocean
  acidification.
\newblock {\em Nature communications}, 10(1):1--10.

\bibitem[Bolotov et~al., 2017]{bolotov2017}
Bolotov, I.~N., Kondakov, A.~V., Vikhrev, I.~V., Aksenova, O.~V., Bespalaya,
  Y.~V., Gofarov, M.~Y., Kolosova, Y.~S., Konopleva, E.~S., Spitsyn, V.~M.,
  Tanmuangpak, K., et~al. (2017).
\newblock Ancient river inference explains exceptional oriental freshwater
  mussel radiations.
\newblock {\em Scientific Reports}, 7(1):1--14.

\bibitem[Bouckaert, 2010]{Bouckaert2010}
Bouckaert, R.~R. (2010).
\newblock Densitree: making sense of sets of phylogenetic trees.
\newblock {\em Bioinformatics}, 26(10):1372--1373.

\bibitem[Breton et~al., 2007]{Breton2007}
Breton, S., Beaupre, H.~D., Stewart, D.~T., Hoeh, W.~R., and Blier, P.~U.
  (2007).
\newblock The unusual system of doubly uniparental inheritance of mt{DNA}:
  isn't one enough?
\newblock {\em Trends in genetics}, 23(9):465--474.

\bibitem[Breton et~al., 2011]{Breton2011}
Breton, S., Stewart, D.~T., Shepardson, S., Trdan, R.~J., Bogan, A.~E.,
  Chapman, E.~G., Ruminas, A.~J., Piontkivska, H., and Hoeh, W.~R. (2011).
\newblock Novel protein genes in animal mt{DNA}: a new sex determination system
  in freshwater mussels (bivalvia: Unionoida)?
\newblock {\em Molecular Biology and Evolution}, 28(5):1645--1659.

\bibitem[Bril et~al., 2017]{bril2017}
Bril, J.~S., Langenfeld, K., Just, C.~L., Spak, S.~N., and Newton, T.~J.
  (2017).
\newblock Simulated mussel mortality thresholds as a function of mussel biomass
  and nutrient loading.
\newblock {\em PeerJ}, 5:e2838.

\bibitem[Cahn, 1936]{Cahn1936}
Cahn, A.~R. (1936).
\newblock {\em The Molluscan Fauna of the Clinch River Below Norris Dam Upon
  the Completion of that Structure}.
\newblock Tennessee Valley Authority.

\bibitem[Campbell et~al., 2005]{Campbell2005}
Campbell, D.~C., Serb, J.~M., Buhay, J.~E., Roe, K.~J., Minton, R.~L., and
  Lydeard, C. (2005).
\newblock Phylogeny of {N}orth {A}merican amblemines (\emph{Bivalvia,
  Unionoida}): prodigious polyphyly proves pervasive across genera.
\newblock {\em Invertebrate Biology}, 124(2):131--164.

\bibitem[Capella-Guti{\'e}rrez et~al., 2009]{Capella2009}
Capella-Guti{\'e}rrez, S., Silla-Mart{\'\i}nez, J.~M., and Gabald{\'o}n, T.
  (2009).
\newblock trimal: a tool for automated alignment trimming in large-scale
  phylogenetic analyses.
\newblock {\em Bioinformatics}, 25(15):1972--1973.

\bibitem[Chu et~al., 2016]{Chu2016}
Chu, C., Nielsen, R., and Wu, Y. (2016).
\newblock {REP}denovo: inferring \emph{de novo} repeat motifs from short
  sequence reads.
\newblock {\em PloS {O}ne}, 11(3):e0150719.

\bibitem[Conant and Wolfe, 2008]{conant2008}
Conant, G.~C. and Wolfe, K.~H. (2008).
\newblock Turning a hobby into a job: how duplicated genes find new functions.
\newblock {\em Nature Reviews Genetics}, 9(12):938--950.

\bibitem[Consortium et~al., 2001]{Lander2001}
Consortium, I. H. G.~S. et~al. (2001).
\newblock Initial sequencing and analysis of the human genome.
\newblock {\em Nature}, 409:860--921.

\bibitem[Cornman et~al., 2014]{Cornman2014}
Cornman, R.~S., Robertson, L.~S., Galbraith, H., and Blakeslee, C. (2014).
\newblock Transcriptomic analysis of the mussel \emph{{E}lliptio complanata}
  identifies candidate stress-response genes and an abundance of novel or
  noncoding transcripts.
\newblock {\em PLoS One}, 9(11):e112420.

\bibitem[Degnan and Rosenberg, 2009]{Degnan2009}
Degnan, J.~H. and Rosenberg, N.~A. (2009).
\newblock Gene tree discordance, phylogenetic inference and the multispecies
  coalescent.
\newblock {\em Trends in ecology \& evolution}, 24(6):332--340.

\bibitem[Demuth et~al., 2006]{Demuth2006}
Demuth, J.~P., De~Bie, T., Stajich, J.~E., Cristianini, N., and Hahn, M.~W.
  (2006).
\newblock The evolution of mammalian gene families.
\newblock {\em PloS One}, 1(1):e85.

\bibitem[Edgar, 2004]{Edgar2004}
Edgar, R.~C. (2004).
\newblock {MUSCLE}: multiple sequence alignment with high accuracy and high
  throughput.
\newblock {\em Nucleic {A}cids {R}esearch}, 32(5):1792--1797.

\bibitem[El-Gebali et~al., 2019]{pfam}
El-Gebali, S., Mistry, J., Bateman, A., Eddy, S.~R., Luciani, A., Potter,
  S.~C., Qureshi, M., Richardson, L.~J., Salazar, G.~A., Smart, A., et~al.
  (2019).
\newblock The pfam protein families database in 2019.
\newblock {\em Nucleic {A}cids {R}esearch}, 47(D1):D427--D432.

\bibitem[\emph{{D}rosophila} 12~Genomes~Consortium et~al., 2007]{12Genomes}
\emph{{D}rosophila} 12~Genomes~Consortium et~al. (2007).
\newblock Evolution of genes and genomes on the \emph{{D}rosophila} phylogeny.
\newblock {\em Nature}, 450(7167):203.

\bibitem[Feschotte, 2008]{Feschotte2008}
Feschotte, C. (2008).
\newblock Transposable elements and the evolution of regulatory networks.
\newblock {\em Nature Reviews Genetics}, 9(5):397.

\bibitem[Garner and McGregor, 2001]{Garner2001}
Garner, J. and McGregor, S. (2001).
\newblock Current status of freshwater mussels ({U}nionidae,
  {M}argaritiferidae) in the {M}uscle {S}hoals area of {T}ennessee {R}iver in
  {A}labama ({M}uscle {S}hoals revisited again).
\newblock {\em American Malacological Bulletin}, 16(1-2):155--170.

\bibitem[Gillespie, 1991]{Gillespie1991}
Gillespie, J.~H. (1991).
\newblock {\em The Causes of Molecular Evolution}, chapter~5, page 232.
\newblock Oxford University Press.

\bibitem[Gillespie, 2004]{Gillespie2004}
Gillespie, J.~H. (2004).
\newblock {\em Population genetics: a concise guide}.
\newblock JHU Press.

\bibitem[Goffeau et~al., 1996]{goffeau1996}
Goffeau, A., Barrell, B.~G., Bussey, H., Davis, R.~W., Dujon, B., Feldmann, H.,
  Galibert, F., Hoheisel, J.~D., Jacq, C., Johnston, M., et~al. (1996).
\newblock Life with 6000 genes.
\newblock {\em Science}, 274(5287):546--567.

\bibitem[Goldman and Yang, 1994]{Goldman1994}
Goldman, N. and Yang, Z. (1994).
\newblock A codon-based model of nucleotide substitution for protein-coding
  {DNA} sequences.
\newblock {\em Molecular Biology and Evolution}, 11(5):725--736.

\bibitem[Gonz{\'a}lez et~al., 2015]{Gonzalez2015}
Gonz{\'a}lez, V.~L., Andrade, S.~C., Bieler, R., Collins, T.~M., Dunn, C.~W.,
  Mikkelsen, P.~M., Taylor, J.~D., and Giribet, G. (2015).
\newblock A phylogenetic backbone for bivalvia: an {RNA}-seq approach.
\newblock {\em Proceedings of the Royal Society B: Biological Sciences},
  282(1801):20142332.

\bibitem[Haag, 2012]{haag2012}
Haag, W.~R. (2012).
\newblock {\em {N}orth {A}merican freshwater mussels: natural history, ecology,
  and conservation}.
\newblock Cambridge University Press.

\bibitem[Haag and Rypel, 2011]{haag2011}
Haag, W.~R. and Rypel, A.~L. (2011).
\newblock Growth and longevity in freshwater mussels: evolutionary and
  conservation implications.
\newblock {\em Biological Reviews}, 86(1):225--247.

\bibitem[Haag and Williams, 2014]{Haag2014}
Haag, W.~R. and Williams, J.~D. (2014).
\newblock Biodiversity on the brink: an assessment of conservation strategies
  for {N}orth {A}merican freshwater mussels.
\newblock {\em Hydrobiologia}, 735(1):45--60.

\bibitem[Haggerty et~al., 2005]{Haggerty2005}
Haggerty, T.~M., Garner, J.~T., and Rogers, R.~L. (2005).
\newblock Reproductive phenology in \emph{Megalonaias nervosa ({B}ivalvia:
  {U}nionidae)} in {W}heeler {R}eservoir, {T}ennessee {R}iver, {A}labama,
  {USA}.
\newblock {\em Hydrobiologia}, 539(1):131--136.

\bibitem[Hahn, 2009]{Hahn2009}
Hahn, M.~W. (2009).
\newblock Distinguishing among evolutionary models for the maintenance of gene
  duplicates.
\newblock {\em The Journal of Heredity}, 100(5):605--617.

\bibitem[Hahn et~al., 2007]{TripleHan}
Hahn, M.~W., Han, M.~V., and Han, S.-G. (2007).
\newblock Gene family evolution across 12 \emph{{D}rosophila} genomes.
\newblock {\em PLoS Genetics}, 3(11):e197.

\bibitem[Han et~al., 2009]{Han2009}
Han, M.~V., Demuth, J.~P., McGrath, C.~L., Casola, C., and Hahn, M.~W. (2009).
\newblock Adaptive evolution of young gene duplicates in mammals.
\newblock {\em Genome Research}, 19(5):859--867.

\bibitem[Hermisson and Pennings, 2005]{Hermisson2005}
Hermisson, J. and Pennings, P.~S. (2005).
\newblock Soft sweeps: molecular population genetics of adaptation from
  standing genetic variation.
\newblock {\em Genetics}, 169(4):2335--2352.

\bibitem[Inoue et~al., 2013]{Inoue2013}
Inoue, K., Hayes, D.~M., Harris, J.~L., and Christian, A.~D. (2013).
\newblock Phylogenetic and morphometric analyses reveal ecophenotypic
  plasticity in freshwater mussels \emph{{O}bovaria jacksoniana} and
  \emph{{V}illosa arkansasensis} ({B}ivalvia: {U}nionidae).
\newblock {\em Ecology and Evolution}, 3(8):2670--2683.

\bibitem[Isom et~al., 1973]{Isom1973}
Isom, B.~G., Yokley, P., and Gooch, C.~H. (1973).
\newblock Mussels of the {E}lk {R}iver {B}asin in {A}labama and
  {T}ennessee-1965-1967.
\newblock {\em American Midland Naturalist}, pages 437--442.

\bibitem[Johnson, 2002]{Johnson2002}
Johnson, C.~N. (2002).
\newblock Determinants of loss of mammal species during the late quaternary
  `megafauna' extinctions: life history and ecology, but not body size.
\newblock {\em Proceedings of the Royal Society of London. Series B: Biological
  Sciences}, 269(1506):2221--2227.

\bibitem[Johnson et~al., 2012]{Johnson2012}
Johnson, J.~A., Wisniewski, J.~M., Fritts, A.~K., and Bringolf, R.~B. (2012).
\newblock Host identification and glochidia morphology of freshwater mussels
  from the {A}ltamaha {R}iver basin.
\newblock {\em Southeastern Naturalist}, 11(4):733--746.

\bibitem[Johnson et~al., 2018]{Johnson2018}
Johnson, N.~A., Smith, C.~H., Pfeiffer, J.~M., Randklev, C.~R., Williams,
  J.~D., and Austin, J.~D. (2018).
\newblock Integrative taxonomy resolves taxonomic uncertainty for freshwater
  mussels being considered for protection under the us endangered species act.
\newblock {\em Scientific reports}, 8(1):1--16.

\bibitem[Jurka et~al., 2005]{jurka2005}
Jurka, J., Kapitonov, V.~V., Pavlicek, A., Klonowski, P., Kohany, O., and
  Walichiewicz, J. (2005).
\newblock Rep{B}ase update, a database of eukaryotic repetitive elements.
\newblock {\em Cytogenetic and Genome Research}, 110(1-4):462--467.

\bibitem[Karasov et~al., 2010]{Karasov2010}
Karasov, T., Messer, P.~W., and Petrov, D.~A. (2010).
\newblock Evidence that adaptation in \emph{{D}rosophila} is not limited by
  mutation at single sites.
\newblock {\em PLoS Genetics}, 6(6):e1000924.

\bibitem[Kent, 2002]{kent2002}
Kent, W.~J. (2002).
\newblock {BLAT} the {BLAST}-like alignment tool.
\newblock {\em Genome Research}, 12(4):656--664.

\bibitem[Kumar et~al., 2017]{TimeTree}
Kumar, S., Stecher, G., Suleski, M., and Hedges, S.~B. (2017).
\newblock Timetree: a resource for timelines, timetrees, and divergence times.
\newblock {\em Molecular Biology and Evolution}, 34(7):1812--1819.

\bibitem[Lake et~al., 2000]{Lake2000}
Lake, P.~S., Palmer, M.~A., Biro, P., Cole, J., Covich, A.~P., Dahm, C.,
  Gibert, J., Goedkoop, W., Martens, K., and Verhoeven, J. (2000).
\newblock Global change and the biodiversity of freshwater ecosystems: Impacts
  on linkages between above-sediment and sediment biota: All forms of
  anthropogenic disturbance---changes in land use, biogeochemical processes, or
  biotic addition or loss---not only damage the biota of freshwater sediments
  but also disrupt the linkages between above-sediment and sediment-dwelling
  biota.
\newblock {\em BioScience}, 50(12):1099--1107.

\bibitem[Larkin et~al., 2007]{clustalwx}
Larkin, M.~A., Blackshields, G., Brown, N.~P., Chenna, R., McGettigan, P.~A.,
  McWilliam, H., Valentin, F., Wallace, I.~M., Wilm, A., Lopez, R., et~al.
  (2007).
\newblock {C}lustal {W} and {C}lustal {X} version 2.0.
\newblock {\em bioinformatics}, 23(21):2947--2948.

\bibitem[Laudencia-Chingcuanco and Fowler, 2012]{laudencia2012}
Laudencia-Chingcuanco, D. and Fowler, D.~B. (2012).
\newblock Genotype-dependent burst of transposable element expression in crowns
  of hexaploid wheat (\emph{{T}riticum aestivum {L.}}) during cold acclimation.
\newblock {\em Comparative and Functional Genomics}, 2012.

\bibitem[Li and Durbin, 2009]{BWA}
Li, H. and Durbin, R. (2009).
\newblock Fast and accurate short read alignment with {B}urrows--{W}heeler
  transform.
\newblock {\em Bioinformatics}, 25(14):1754--1760.

\bibitem[Li and Durbin, 2011]{PSMC}
Li, H. and Durbin, R. (2011).
\newblock Inference of human population history from individual whole-genome
  sequences.
\newblock {\em Nature}, 475(7357):493.

\bibitem[Liu et~al., 1996]{Liu1996}
Liu, H.-P., Mitton, J.~B., and Wu, S.-K. (1996).
\newblock Paternal mitochondrial {DNA} differentiation far exceeds maternal
  mitochondrial {DNA} and allozyme differentiation in the freshwater mussel,
  \emph{Anodonta grandis grandis}.
\newblock {\em Evolution}, 50(2):952--957.

\bibitem[Loganathan et~al., 1998]{loganathan1998}
Loganathan, B., Neale, J., Sickel, J., Sajwan, K., and Owen, D. (1998).
\newblock Persistent organochlorine concentrations in sediment and mussel
  tissues from the lowermost {T}ennessee {R}iver and {K}entucky lake, {USA}.
\newblock {\em Organohalogen Compounds}, 39:121--124.

\bibitem[Lopes-Lima et~al., 2018]{Lopes2018}
Lopes-Lima, M., Burlakova, L.~E., Karatayev, A.~Y., Mehler, K., Seddon, M., and
  Sousa, R. (2018).
\newblock Conservation of freshwater bivalves at the global scale: diversity,
  threats and research needs.
\newblock {\em Hydrobiologia}, 810(1):1--14.

\bibitem[Lopez et~al., 2020]{Lopez2020}
Lopez, J.~W., Parr, T.~B., Allen, D.~C., and Vaughn, C.~C. (2020).
\newblock Animal aggregations promote emergent aquatic plant production at the
  aquatic-terrestrial interface.
\newblock {\em Ecology}, page e03126.

\bibitem[Luo et~al., 2014]{Luo2014}
Luo, Y., Li, C., Landis, A.~G., Wang, G., Stoeckel, J., and Peatman, E. (2014).
\newblock Transcriptomic profiling of differential responses to drought in two
  freshwater mussel species, the giant floater \emph{{P}yganodon grandis} and
  the pondhorn \emph{{U}niomerus tetralasmus}.
\newblock {\em PLoS One}, 9(2):e89481.

\bibitem[Lynch, 2007]{LynchBook}
Lynch, M. (2007).
\newblock {\em The Origins of Genome Architecture}, page 494.
\newblock Sinauer Associates, Sunderland, Mass.

\bibitem[Lynch and Conery, 2000]{Lynch2000}
Lynch, M. and Conery, J.~S. (2000).
\newblock The evolutionary fate and consequences of duplicate genes.
\newblock {\em Science}, 290(5494):1151--1155.

\bibitem[Lynch et~al., 2011]{Lynch2011}
Lynch, V.~J., Leclerc, R.~D., May, G., and Wagner, G.~P. (2011).
\newblock Transposon-mediated rewiring of gene regulatory networks contributed
  to the evolution of pregnancy in mammals.
\newblock {\em Nature Genetics}, 43(11):1154.

\bibitem[MacManes, 2018]{ORP}
MacManes, M.~D. (2018).
\newblock The {O}yster {R}iver {P}rotocol: a multi-assembler and kmer approach
  for \emph{de novo} transcriptome assembly.
\newblock {\em PeerJ}, 6:e5428.

\bibitem[Marmeisse et~al., 2013]{marmeisse2013}
Marmeisse, R., Nehls, U., {\"O}pik, M., Selosse, M.-A., and Pringle, A. (2013).
\newblock Bridging mycorrhizal genomics, metagenomics and forest ecology.
\newblock {\em New Phytologist}, 198(2):343--346.

\bibitem[Maynard~Smith, 1971]{Maynard1971}
Maynard~Smith, J. (1971).
\newblock What use is sex?
\newblock {\em Journal of Theoretical Biology}, 30(2):319--335.

\bibitem[McKenna et~al., 2010]{gatk}
McKenna, A., Hanna, M., Banks, E., Sivachenko, A., Cibulskis, K., Kernytsky,
  A., Garimella, K., Altshuler, D., Gabriel, S., Daly, M., et~al. (2010).
\newblock The genome analysis toolkit: a mapreduce framework for analyzing
  next-generation {DNA} sequencing data.
\newblock {\em Genome Research}, 20(9):1297--1303.

\bibitem[Milam et~al., 2005]{milam2005}
Milam, C., Farris, J., Dwyer, F., and Hardesty, D. (2005).
\newblock Acute toxicity of six freshwater mussel species (glochidia) to six
  chemicals: Implications for daphnids and \emph{{U}tterbackia imbecillis} as
  surrogates for protection of freshwater mussels ({U}nionidae).
\newblock {\em Archives of Environmental Contamination and Toxicology},
  48(2):166--173.

\bibitem[Moore, 1995]{moore1995}
Moore, W.~S. (1995).
\newblock Inferring phylogenies from mt{DNA} variation: mitochondrial-gene
  trees versus nuclear-gene trees.
\newblock {\em Evolution}, 49(4):718--726.

\bibitem[Nei and Gojobori, 1986]{Nei1986}
Nei, M. and Gojobori, T. (1986).
\newblock Simple methods for estimating the numbers of synonymous and
  nonsynonymous nucleotide substitutions.
\newblock {\em Molecular biology and evolution}, 3(5):418--426.

\bibitem[Nelson et~al., 2004]{Nelson2004}
Nelson, D.~R., Zeldin, D.~C., Hoffman, S.~M., Maltais, L.~J., Wain, H.~M., and
  Nebert, D.~W. (2004).
\newblock Comparison of cytochrome {P}450 ({CYP}) genes from the mouse and
  human genomes, including nomenclature recommendations for genes, pseudogenes
  and alternative-splice variants.
\newblock {\em Pharmacogenetics and Genomics}, 14(1):1--18.

\bibitem[Ohno, 1970]{Ohno1970}
Ohno, S. (1970).
\newblock {\em Evolution by Gene Duplication}.
\newblock Springer-Verlag, Berlin, New York.

\bibitem[Ohta, 1973]{Ohta1973}
Ohta, T. (1973).
\newblock Slightly deleterious mutant substitutions in evolution.
\newblock {\em Nature}, 246(5428):96--98.

\bibitem[Oliver and Greene, 2009]{oliver2009}
Oliver, K.~R. and Greene, W.~K. (2009).
\newblock Transposable elements: powerful facilitators of evolution.
\newblock {\em Bioessays}, 31(7):703--714.

\bibitem[Orr, 2005]{Orr2005}
Orr, H.~A. (2005).
\newblock The genetic theory of adaptation: a brief history.
\newblock {\em Nature Reviews Genetics}, 6(2):119--127.

\bibitem[Ortmann, 1924]{Ortmann1924}
Ortmann, A. (1924).
\newblock {M}ussel {S}hoals.
\newblock {\em Science}, 60(1564):565--566.

\bibitem[Pereira et~al., 2011]{Pereira2011}
Pereira, J.~C., Chaves, R., Bastos, E., Leit{\~a}o, A., and Guedes-Pinto, H.
  (2011).
\newblock An efficient method for genomic {DNA} extraction from different
  molluscs species.
\newblock {\em International journal of molecular sciences}, 12(11):8086--8095.

\bibitem[Pertea and Pertea, 2020]{gffread}
Pertea, G. and Pertea, M. (2020).
\newblock Gff utilities: {GFFR}ead and {GFFC}ompare.
\newblock {\em F1000Research}, 9(304):304.

\bibitem[Pfeiffer et~al., 2019]{Pfeiffer2019}
Pfeiffer, J.~M., Breinholt, J.~W., and Page, L.~M. (2019).
\newblock Unioverse: A phylogenomic resource for reconstructing the evolution
  of freshwater mussels (bivalvia, unionoida).
\newblock {\em Molecular phylogenetics and evolution}, 137:114--126.

\bibitem[Pfeiffer et~al., 2016]{Pfeiffer2016}
Pfeiffer, J.~M., Johnson, N.~A., Randklev, C.~R., Howells, R.~G., and Williams,
  J.~D. (2016).
\newblock Generic reclassification and species boundaries in the rediscovered
  freshwater mussel \emph{`Quadrula' mitchelli} (simpson in dall, 1896).
\newblock {\em Conservation Genetics}, 17(2):279--292.

\bibitem[Pfeiffer et~al., 2018]{Pfeiffer2018}
Pfeiffer, J.~M., Sharpe, A.~E., Johnson, N.~A., Emery, K.~F., and Page, L.~M.
  (2018).
\newblock Molecular phylogeny of the {N}earctic and {M}esoamerican freshwater
  mussel genus \emph{{M}egalonaias}.
\newblock {\em Hydrobiologia}, 811(1):139--151.

\bibitem[Pogson and Zouros, 1994]{scallop}
Pogson, G.~H. and Zouros, E. (1994).
\newblock Allozyme and rflp heterozygosities as correlates of growth rate in
  the scallop \emph{{P}lacopecten magellanicus}: a test of the associative
  overdominance hypothesis.
\newblock {\em Genetics}, 137(1):221--231.

\bibitem[Price et~al., 2005]{RepeatScout}
Price, A.~L., Jones, N.~C., and Pevzner, P.~A. (2005).
\newblock \emph{De novo} identification of repeat families in large genomes.
\newblock {\em Bioinformatics}, 21(suppl\_1):i351--i358.

\bibitem[Quevillon et~al., 2005]{quevillon2005}
Quevillon, E., Silventoinen, V., Pillai, S., Harte, N., Mulder, N., Apweiler,
  R., and Lopez, R. (2005).
\newblock {I}nter{P}ro{S}can: protein domains identifier.
\newblock {\em Nucleic Acids Research}, 33(suppl\_2):W116--W120.

\bibitem[Renaut et~al., 2018]{Renaut2018}
Renaut, S., Guerra, D., Hoeh, W.~R., Stewart, D.~T., Bogan, A.~E., Ghiselli,
  F., Milani, L., Passamonti, M., and Breton, S. (2018).
\newblock Genome survey of the freshwater mussel \emph{Venustaconcha
  ellipsiformis (Bivalvia: Unionida)} using a hybrid \emph{de novo} assembly
  approach.
\newblock {\em Genome Biology and Evolution}, 10(7):1637--1646.

\bibitem[Rogers et~al., 2009]{Rogers2009}
Rogers, R.~L., Bedford, T., and Hartl, D.~L. (2009).
\newblock Formation and longevity of chimeric and duplicate genes in
  \emph{{D}rosophila melanogaster}.
\newblock {\em Genetics}, 181(1):313--322.

\bibitem[Rogers et~al., 2014]{Rogers2014}
Rogers, R.~L., Cridland, J.~M., Shao, L., Hu, T.~T., Andolfatto, P., and
  Thornton, K.~R. (2014).
\newblock Landscape of standing variation for tandem duplications in
  \emph{{D}rosophila yakuba} and \emph{{D}rosophila simulans}.
\newblock {\em Molecular Biology and Evolution}, 31:1750--1766.

\bibitem[Rogers et~al., 2017]{Rogers2017Exp}
Rogers, R.~L., Shao, L., and Thornton, K.~R. (2017).
\newblock Tandem duplications lead to novel expression patterns through exon
  shuffling in \emph{{D}rosophila yakuba}.
\newblock {\em PLoS Genetics}, 13(5):e1006795.

\bibitem[Sayyari and Mirarab, 2016]{Sayyari2016}
Sayyari, E. and Mirarab, S. (2016).
\newblock Fast coalescent-based computation of local branch support from
  quartet frequencies.
\newblock {\em Molecular biology and evolution}, 33(7):1654--1668.

\bibitem[Schaack et~al., 2010]{Schaack2010}
Schaack, S., Gilbert, C., and Feschotte, C. (2010).
\newblock Promiscuous {DNA}: horizontal transfer of transposable elements and
  why it matters for eukaryotic evolution.
\newblock {\em Trends in Ecology \& Evolution}, 25(9):537--546.

\bibitem[Schliep, 2011]{Schliep2011}
Schliep, K.~P. (2011).
\newblock phangorn: phylogenetic analysis in {R}.
\newblock {\em Bioinformatics}, 27(4):592--593.

\bibitem[Schmidt et~al., 2017]{Schmidt2017}
Schmidt, J.~M., Battlay, P., Gledhill-Smith, R.~S., Good, R.~T., Lumb, C.,
  Fournier-Level, A., and Robin, C. (2017).
\newblock Insights into {DDT} resistance from the \emph{{D}rosophila
  melanogaster} genetic reference panel.
\newblock {\em Genetics}, 207(3):1181--1193.

\bibitem[Schmidt et~al., 2010]{schmidt2010}
Schmidt, J.~M., Good, R.~T., Appleton, B., Sherrard, J., Raymant, G.~C.,
  Bogwitz, M.~R., Martin, J., Daborn, P.~J., Goddard, M.~E., Batterham, P.,
  et~al. (2010).
\newblock Copy number variation and transposable elements feature in recent,
  ongoing adaptation at the \emph{Cyp6g1} locus.
\newblock {\em PLoS Genet}, 6(6):e1000998.

\bibitem[Sim{\~a}o et~al., 2015]{simao2015}
Sim{\~a}o, F.~A., Waterhouse, R.~M., Ioannidis, P., Kriventseva, E.~V., and
  Zdobnov, E.~M. (2015).
\newblock {BUSCO}: assessing genome assembly and annotation completeness with
  single-copy orthologs.
\newblock {\em Bioinformatics}, 31(19):3210--3212.

\bibitem[Smith et~al., 2020]{Smith2020}
Smith, C.~H., Pfeiffer, J.~M., and Johnson, N.~A. (2020).
\newblock Comparative phylogenomics reveal complex evolution of life history
  strategies in a clade of bivalves with parasitic larvae (bivalvia: Unionoida:
  Ambleminae).
\newblock {\em Cladistics}, 36(5):505--520.

\bibitem[S{\o}rensen et~al., 2003]{Sorensen2003}
S{\o}rensen, J.~G., Kristensen, T.~N., and Loeschcke, V. (2003).
\newblock The evolutionary and ecological role of heat shock proteins.
\newblock {\em Ecology Letters}, 6(11):1025--1037.

\bibitem[Stamatakis, 2006]{Stamatakis2006}
Stamatakis, A. (2006).
\newblock {RAxML-VI-HPC}: maximum likelihood-based phylogenetic analyses with
  thousands of taxa and mixed models.
\newblock {\em Bioinformatics}, 22(21):2688--2690.

\bibitem[Stamatakis, 2014]{Stamatakis2014}
Stamatakis, A. (2014).
\newblock {RAxML} version 8: a tool for phylogenetic analysis and post-analysis
  of large phylogenies.
\newblock {\em Bioinformatics}, 30(9):1312--1313.

\bibitem[Stanke et~al., 2008]{stanke2008}
Stanke, M., Diekhans, M., Baertsch, R., and Haussler, D. (2008).
\newblock Using native and syntenically mapped c{DNA} alignments to improve
  \emph{de novo} gene finding.
\newblock {\em Bioinformatics}, 24(5):637--644.

\bibitem[Stanke and Waack, 2003]{stanke2003}
Stanke, M. and Waack, S. (2003).
\newblock Gene prediction with a hidden {M}arkov model and a new intron
  submodel.
\newblock {\em Bioinformatics}, 19(suppl\_2):ii215--ii225.

\bibitem[Stearns, 2000]{Stearns2000}
Stearns, S.~C. (2000).
\newblock Life history evolution: successes, limitations, and prospects.
\newblock {\em Naturwissenschaften}, 87(11):476--486.

\bibitem[Strayer et~al., 2004]{Strayer2004}
Strayer, D.~L., Downing, J.~A., Haag, W.~R., King, T.~L., Layzer, J.~B.,
  Newton, T.~J., and Nichols, J.~S. (2004).
\newblock Changing perspectives on pearly mussels, {N}orth {A}merica's most
  imperiled animals.
\newblock {\em BioScience}, 54(5):429--439.

\bibitem[Thompson et~al., 1994]{clustalw}
Thompson, J.~D., Higgins, D.~G., and Gibson, T.~J. (1994).
\newblock {CLUSTAL W}: improving the sensitivity of progressive multiple
  sequence alignment through sequence weighting, position-specific gap
  penalties and weight matrix choice.
\newblock {\em Nucleic Acids Research}, 22(22):4673--4680.

\bibitem[Toews and Brelsford, 2012]{toews2012}
Toews, D.~P. and Brelsford, A. (2012).
\newblock The biogeography of mitochondrial and nuclear discordance in animals.
\newblock {\em Molecular Ecology}, 21(16):3907--3930.

\bibitem[Vaughn, 2018]{Vaughn2018}
Vaughn, C.~C. (2018).
\newblock Ecosystem services provided by freshwater mussels.
\newblock {\em Hydrobiologia}, 810(1):15--27.

\bibitem[Vaughn et~al., 2008]{Vaughn2008}
Vaughn, C.~C., Nichols, S.~J., and Spooner, D.~E. (2008).
\newblock Community and foodweb ecology of freshwater mussels.
\newblock {\em Journal of the North American Benthological Society},
  27(2):409--423.

\bibitem[Wakeley, 2009]{Wakeley2009}
Wakeley, J. (2009).
\newblock {\em Coalescent Theory: An Introduction}, page~97.
\newblock Roberts \& Company Publishers.

\bibitem[Wang et~al., 2012]{Wang2012}
Wang, R., Li, C., Stoeckel, J., Moyer, G., Liu, Z., and Peatman, E. (2012).
\newblock Rapid development of molecular resources for a freshwater mussel,
  \emph{{V}illosa lienosa} ({B}ivalvia: {U}nionidae), using an {RNA}-seq-based
  approach.
\newblock {\em Freshwater Science}, 31(3):695--708.

\bibitem[Waterhouse et~al., 2017]{waterhouse2017}
Waterhouse, R.~M., Seppey, M., Sim{\~a}o, F.~A., Manni, M., Ioannidis, P.,
  Klioutchnikov, G., Kriventseva, E.~V., and Zdobnov, E.~M. (2017).
\newblock {BUSCO} applications from quality assessments to gene prediction and
  phylogenomics.
\newblock {\em Molecular Biology and Evolution}, 35(3):543--548.

\bibitem[Watterson, 1975]{watterson1975}
Watterson, G. (1975).
\newblock On the number of segregating sites in genetical models without
  recombination.
\newblock {\em Theoretical Population Biology}, 7(2):256--276.

\bibitem[Wen et~al., 2017]{Wen2017}
Wen, H.~B., Cao, Z.~M., Hua, D., Xu, P., Ma, X.~Y., Jin, W., Yuan, X.~H., and
  Gu, R.~B. (2017).
\newblock The complete maternally and paternally inherited mitochondrial
  genomes of a freshwater mussel \emph{Potamilus alatus (Bivalvia: Unionidae)}.
\newblock {\em PloS {O}ne}, 12(1):e0169749.

\bibitem[Williams et~al., 2008]{williams2008}
Williams, J.~D., Bogan, A.~E., Garner, J.~T., et~al. (2008).
\newblock {\em Freshwater mussels of {A}labama and the {M}obile basin in
  {G}eorgia, {M}ississippi, and {T}ennessee}.
\newblock University of {A}labama Press.

\bibitem[Williams et~al., 1993]{williams1993}
Williams, J.~D., Warren~Jr, M.~L., Cummings, K.~S., Harris, J.~L., and Neves,
  R.~J. (1993).
\newblock Conservation status of freshwater mussels of the {U}nited {S}tates
  and {C}anada.
\newblock {\em Fisheries}, 18(9):6--22.

\bibitem[Woodside, 2004]{woodside2004}
Woodside, M.~D. (2004).
\newblock {\em Water quality in the lower {T}ennessee {R}iver {B}asin,
  {T}ennessee, {A}labama, {K}entucky, {M}ississippi, and {G}eorgia, 1999-2001},
  volume 1233.
\newblock US Geological Survey.

\bibitem[Woody and Holland-Bartels, 1993]{woody1993}
Woody, C.~A. and Holland-Bartels, L. (1993).
\newblock Reproductive characteristics of a population of the washboard mussel
  \emph{Megalonaias nervosa} ({R}afinesque 1820) in the upper {M}ississippi
  {R}iver.
\newblock {\em Journal of Freshwater Ecology}, 8(1):57--66.

\bibitem[Wright, 1931]{Wright1931}
Wright, S. (1931).
\newblock Evolution in mendelian populations.
\newblock {\em Genetics}, 16(2):97.

\bibitem[Wright, 1951]{Wright1951}
Wright, S. (1951).
\newblock The genetical structure of populations.
\newblock {\em Annals of Eugenics}, 15(1):323--354.

\bibitem[Yang, 1997]{PAML}
Yang, Z. (1997).
\newblock {PAML}: a program package for phylogenetic analysis by maximum
  likelihood.
\newblock {\em Bioinformatics}, 13(5):555--556.

\bibitem[Zhang et~al., 2018]{Zhang2018}
Zhang, C., Rabiee, M., Sayyari, E., and Mirarab, S. (2018).
\newblock {ASTRAL-III}: polynomial time species tree reconstruction from
  partially resolved gene trees.
\newblock {\em {BMC} bioinformatics}, 19(6):153.

\bibitem[Zhang et~al., 2012]{oyster2012}
Zhang, G., Fang, X., Guo, X., Li, L., Luo, R., Xu, F., Yang, P., Zhang, L.,
  Wang, X., Qi, H., et~al. (2012).
\newblock The oyster genome reveals stress adaptation and complexity of shell
  formation.
\newblock {\em Nature}, 490(7418):49--54.

\bibitem[Zimin et~al., 2013]{masurca}
Zimin, A.~V., Mar{\c{c}}ais, G., Puiu, D., Roberts, M., Salzberg, S.~L., and
  Yorke, J.~A. (2013).
\newblock The masurca genome assembler.
\newblock {\em Bioinformatics}, 29(21):2669--2677.

\end{thebibliography}
\bibliographystyle{apalike}

\clearpage
\begin{table}
\begin{center}
\caption{Population genetic parameters of three species of bivalves under moderate selection.}
\begin{tabular}{lllll}
\hline
Species & $\theta=4N_e\mu$ & $N_e$& $P_{sgv}$ & $T_e$ (gens) \\ 
\hline
\Mnerv & 0.00770 &388,500 & 0.0672 &12,870 \\
\Velli & 0.00600 & 300,000&0.0509 & 16,667\\
\Ehope & 0.00576 & 288,000 &0.0486  & 56,180\\
\hline
assumes $s=0.01$ & && \\
\end{tabular}
\label{Pgen}
\end{center}
\end{table}

\clearpage

\clearpage

\clearpage

\clearpage

\clearpage
\clearpage

\clearpage

\clearpage
\clearpage

\clearpage
\subsubsection*{Supporting Information}

\renewcommand{\thefigure}{S\arabic{figure}}
\renewcommand{\thetable}{S\arabic{table}}
\setcounter{figure}{0}
\setcounter{table}{0}

\subsubsection*{Transcriptome annotation for other \Unio}
Comparisons of gene content across species are complicated by varying assembly quality and limited genome availability for \Unio.  Transcriptome data for \EcrasFull {} originally collected for other purposes were incidentally available.  Fresh gonad tissue was dissected for a single individual from the Apalachicola River Basin (Georgia, USA) in 2013.  Gonad tissue was fixed in RNAlater and prepped using a Trizol RNA isolation reaction. The reactions were quantified on a qubit at 1 ng/$\mu$l RNA. RNA was prepared for RNA sequencing at the Georgia Genomic Core Facility.  Sequencing was performed on a MiSeq with 150 bp paired end reads.  

Previously published transcriptome sequences were available in the SRA for for \VelliFull {} \citep{Renaut2018}, \Ecomp {} \citep{Cornman2014}, \Llien {} \citep{Wang2012}, \Utet {} \citep{Luo2014}, \Aplic {} (Table \ref{SRANos}).  To offer a comparison for gene content in \Mnerv {} that could be verified independently from genome assembly, we \emph{de novo} assembled and annotated transcriptomes for these 5 species and the highest coverage transcriptome from mucosal palps for \Mnerv.   Transcriptomes were assembled using the Oyster River Protocol \citep{ORP} then translated into six open reading frames.    Translated proteins were functionally annotated using Interproscan \citep{quevillon2005}, which identifies conserved domains in translated protein sequences.  These results offer comparable annotations for all species, though with less stringent criteria than employed when mapping to genome assemblies for official annotations.  The analysis offers comparisons that are independent from genome annotation.  Number of transcripts identified could be affected by isoform variability, expression level differences, collapsing highly similar duplicate copies into a single transcript, and tissue-specific expression in different samples, factors that the Oyster River Protocol is designed to mitigate \citep{ORP}.   

The \Mnerv {} transcriptome shows 72.9\% complete, 9.5\% missing BUSCOs. \Ecras {} has 78\% complete, only 4.2\% missing BUSCOs. \Velli {} shows 96\% complete, 0.3\% missing BUSCOs. These transcriptome assemblies suggest a slight bias toward identifying full transcripts in \Velli {} but should be more directly comparable than results from genome assemblies of varying qualities with differing methods of annotation.   Quality control statistics suggest low assembly quality for the \Lcar {} transcriptome with only 37.3\% complete BUSCOs in spite of high sequence quality.  This species was not included in phylogenetic analysis or assays of gene content, although 1:1 orthologs with \Mnerv {} are included in Supplementary Information.

Across the \Unio, there is considerable variation in copy number for \emph{cytochrome P450} genes ranging from 129 copies in \Velli {} and 514 copies in \Llien.  ABC transporters show similar variability, with the highest copy number again in \Llien {} with 350 copies and lowest in \Aplic {} with only 61 copies.  \MnervFull {} shows moderate copy number based on these transcriptome analyses, suggesting that it is not an outlier among \Unio.  We observe striking differences among members of the same genus, with for example 440 cytochromes in \Ecomp {} and 280 in \Ecras.  Glutathione transferases and thioredoxins show less variation in copy number, but with roughly 2-fold higher copy number in \Llien.  The fact that closely related species may vary widely in copy number may suggest dynamic evolution in gene content through gene family expansion and contraction, a possibility that may be addressed more completely with full genome assembly in the future.

\subsubsection*{Phylogenetic placement of \Aplic}
We attempted to place \Aplic {} among the phylogeny, but were able to find no clear consensus placement.  Out of 4178 gene trees with 1:1 orthologs present across all 7 species, 1242 trees out of 4178 cluster \Aplic {} with \emph{Elliptio},  1399 cluster \Aplic {} with \Velli-\Llien, and 205 cluster with \Mnerv-\Utet, and 601 place \Aplic {} external to the split of \Velli-\Llien from \emph{Elliptio}.  It is possible that this species has complex evolutionary processes affecting gene-tree relationships, a prospect deserving of future study with better genetic data.   


\clearpage
 \begin{table}
\begin{center}
\vspace{-0.2in}
\caption{\Mnerv {} reference genome quality}
\label{GenomeStats} 
\begin{tabular}{ lr }
\hline
 contig N50   & 51,552 \\ 
 scaffold N50 & 52,791\\
 max contig length  & 588,638\\
 max scaffold length & 588,638\\
 \hline
Greater than 10 kb & 92\%\\
\hline
\end{tabular}
\end{center}
\vspace{-0.2in}
\end{table}

\clearpage

\begin{table}
\begin{center}
\footnotesize
\caption{Number of Gene Family Members by Functional Class\label{CompVelli}}
\begin{tabular}{lrrr}
\hline
 Function &  \Mnerv  & \Velli  \\
 Cytochrome P450s& 143 & 48 \\
 ABC transporters & 106 &  53 \\
 Hsp70s & 96 & 52 \\
 Opsins & 238 & 5 \\
 Mitochondria eating proteins  & 117 & 6  \\
 Chitin metabolism & 66 & 14 \\
 von Willebrand factors & 138 & 4 \\
\hline
\end{tabular}
\end{center}
\end{table}

\clearpage
\begin{table}
\begin{center}
\caption{SRA Numbers for sequence data \label{SRANos}}
\begin{tabular}{lll}
\hline
\Mnerv &genome & PRJNA646917  \\
& transcriptome & PRJNA646778 \\
\Ehope & genome & PRJNA647325 \\
\Ecras & transcriptome & PRJNA647326 \\
\Ecomp & transcriptome & SRR5136467 \\
 \Utet & transcriptome &  SRR910418 \\ 
 \Llien & transcriptome & SRR354206 \\
 \Velli & genome & SRR6689532 \\
 & & SRR6689533 \\
 & transcriptome & SRR6279374 \\ 
\hline
\end{tabular}
\end{center}
\end{table}

\clearpage

\begin{table}
\begin{center}
\tiny
\caption{Functional classes in \emph{de novo} assembled transcriptomes}
\label{TransCypDetox}
\begin{tabular}{lrrrrrrrr}
Function& \Velli & \Llien & \Ecras & \Ecomp & \Utet & \Mnerv & \Aplic \\
\hline
Cytochrome P450s & 129  & 514  & 280 & 440 & 206 &242 & 181 \\
ABC transporters & 262  &  350 & 243 & 167  &  192 & 122 & 61 \\
Glutathione transferase & 81 & 209& 76 & 108 & 96 & 131 & 93\\
Thioredoxins & 154 & 225 & 76 & 117 & 100 & 114 & 86\\
Hsp70s & 52  & 179 & 313 & 75 & 86 &117 & 103 \\
Mitochondrial eating  & 90 & 145 & 487 & 144 & 191 & 210 & 344 \\
Chitin metabolism & 92 &  248 & 175 & 175 & 72 & 104 & 90 \\
von Willebrand Factors & 130 & 617 &  378 & 313 & 270 & 209 & 175\\
 \hline
\end{tabular}
\end{center}
\end{table}

\clearpage

\begin{center}
\begin{table}
\footnotesize
\caption{Birth-death parameters for gene families}
\label{FamBirthDeath}
\begin{tabular}{lrrrrrr}
Class & Total  & Paralogs & Death rate $\mu$ & $\mu$ fold-change  & Birth rate $\lambda$ & $\lambda$ fold-change \\ 
\hline
All & 49,149 & 6002 & 6.8 & -& 40859  & - \\ 
\hline
Cytochrome P450 & 143 & 88 & 10.2 & 1.5 & 898 & 7.6 \\
Opsin & 238 & 110 & 7.2  & 1.1 & 793& 4.0 \\
Chitin metabolism & 66 & 35 & 6.4 & 0.9 & 224 & 4.0 \\
Mt Eating protein & 117 & 55 & 6.4 & 0.9& 353 & 3.6 \\
Hsp70 & 96 & 61 & 4.6 & 0.7 & 283  & 3.5 \\
VonWillebrand factor & 138 & 52 & 7.8 & 1.1 & 406 & 3.5 \\
ABC transporter & 106 & 28 & 4.3  & 0.6& 122 & 1.4 \\
\hline
\end{tabular}
\end{table}
\end{center}
\clearpage

\begin{table}
\begin{center}
\caption{Interproscan TE Terms \label{TETerms}}
\begin{tabular}{l}
\hline
Integrase \\ 
Reverse transcriptase\\ 
hAT\\ 
Gypsy\\ 
LTR\\  
gag \\ 
Transposase\\ 
RNase H\\ 
DDE superfamily endonuclease\\ 
HNH \\ 
Pox \\
\hline
\end{tabular}
\end{center}
\end{table}

\clearpage
\begin{table}
\begin{center}
\caption{Population genetic parameters of three species of bivalves under strong selection.}
\begin{tabular}{lllll}
\hline
Species & $\theta=4N_e\mu$ & $N_e$& $P_{sgv}$ & $T_e$ (gens) \\ 
\hline
\Mnerv & 0.00770 &388,500 & 0.0837 &1,287 \\
\Velli & 0.00600 & 300,000&0.0649 & 1,667\\
\Ehope & 0.00576 & 288,000 &0.0611  & 5,618\\
\hline
assumes $s=0.1$ & && \\
\end{tabular}
\label{Pgenhigh}
\end{center}
\end{table}

\clearpage
\begin{table}
\begin{center}
\caption{Gene Trees supporting phylogenetic relationships in \Unio}
\begin{tabular}{lrrr}
Species Split & Support & Total & Proportion \\
\hline
Consensus Tree & 2131 & 4475 & 0.4762 \\
\hline
\Mnerv-\Utet & 2966  &  & 0.6628 \\
 \Ecomp-\Ecras & 2964 &   & 0.6623 \\
\Llien-\Velli & 3281 &  &  0.7332 \\
\hline
\Utet-(\Ecomp, \Ecras) & 190 &  & 0.0425\\
\Utet-(\Llien,\Velli) & 302 &  & 0.0675 \\
\Velli-(\Ecomp, \Ecras) &  220 &  & 0.0492 \\
\Llien-(\Ecomp, \Ecras) &  118 &  &  0.0264 \\
\hline

\end{tabular}
\label{GeneTrees}
\end{center}
\end{table}

\clearpage
\begin{figure}
\begin{center}
\vspace{-.35in}

\includegraphics[scale=0.45]{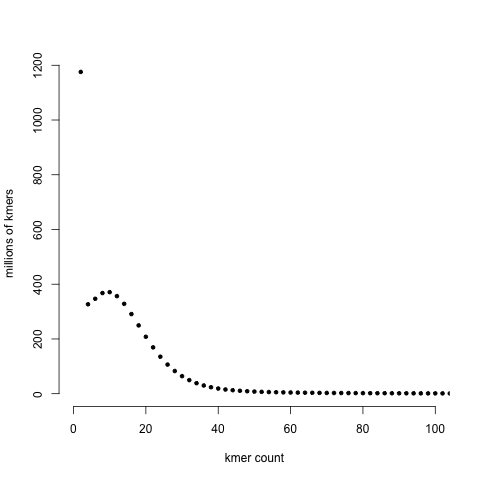}
\end{center}    
\vspace{-.3in}
\caption{\footnotesize{Distribution of kmer counts for one sequencing lane of \Mnerv.  The distribution shows no suggestion of polyploidy.} \label{kmer}}
\vspace{-0.1in}
\end{figure} 

\clearpage
\begin{figure}
\begin{center}
\includegraphics[width=0.7\textwidth]{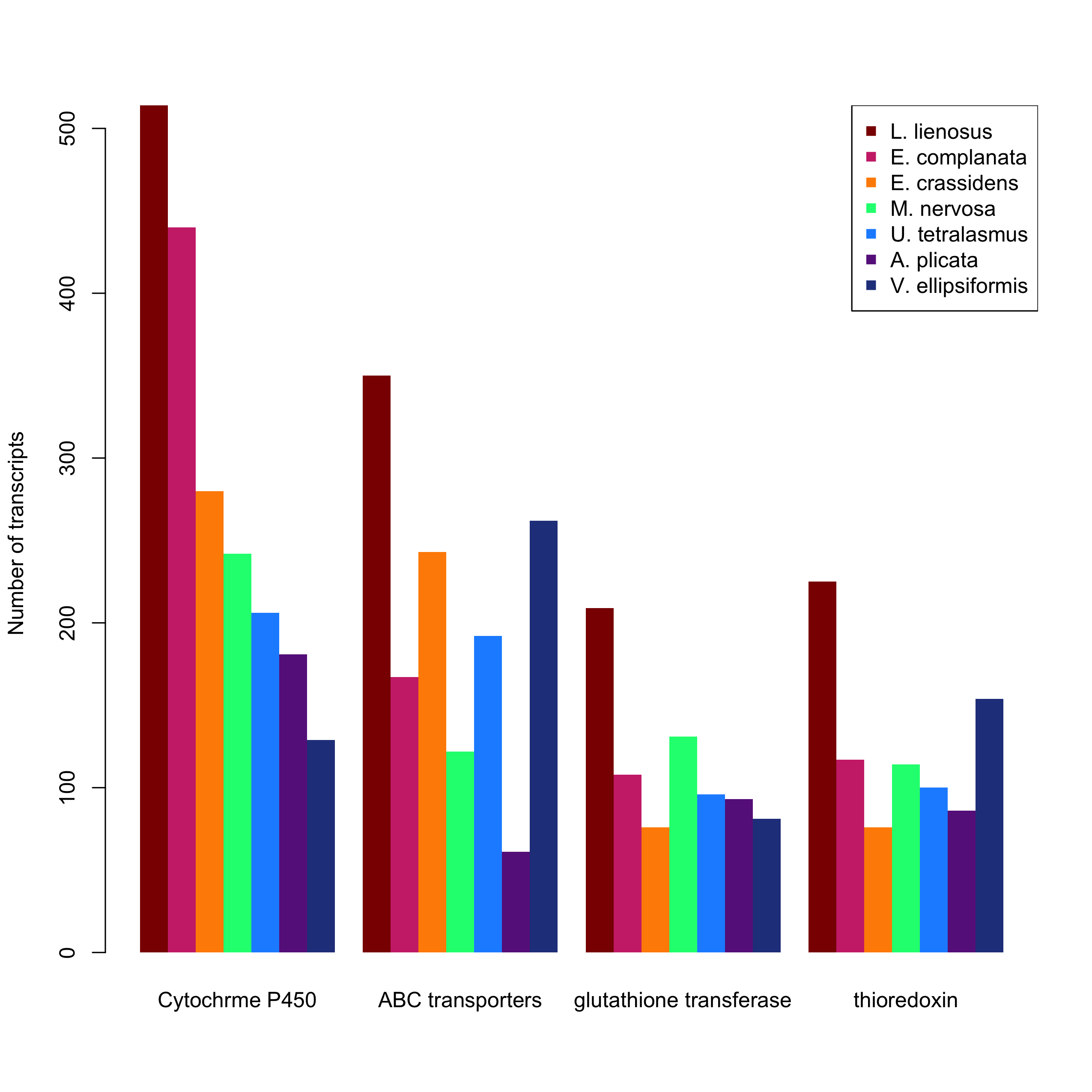}
\caption{Detox genes identified in the transcriptomes of 6 species of \Unio. Detox gene amplification is common in  \Unio, with large detox gene content across all 6 species.  \label{detox2}}
\end{center}

\end{figure}

\end{document}